\documentclass[reprint,english,superscriptaddress,preprintnumbers,amsmath,amssymb,aps,prd]{revtex4-2}

\usepackage[utf8]{inputenc}
\usepackage{diagbox}
\usepackage{orcidlink}
\usepackage{tikz,xcolor,hyperref}
\usepackage{mathtools}
\usepackage{amsfonts}
\usepackage{mathrsfs}
\usepackage{bbm}
\usepackage{slashed}
\usepackage{titlesec}
\usepackage{amssymb}
\usepackage{graphicx}
\usepackage{color}
\usepackage{array}

\usepackage{blindtext}
\usepackage{placeins}
\usepackage{booktabs}
\usepackage{makecell}
\usepackage{epstopdf}
\usepackage[caption=false]{subfig}

\usepackage{xspace}
\usepackage{siunitx}
\usepackage{hyperref}
\usepackage[nameinlink]{cleveref}
\usepackage{appendix}

\usepackage[normalem]{ulem}
\usepackage{xifthen}
\usepackage{xcolor}
\hypersetup{
	colorlinks,
	linkcolor={blue!75!black},
	citecolor={blue!75!black},
	urlcolor={blue!75!black}
}

\setkeys{Gin}{width=0.48\textwidth}

\def\eq#1{(\ref{#1})}

\def\Eq#1{Eq.~(\ref{#1})}

\def\Fig#1{Fig.~\ref{#1}}

\def\Sec#1{Sec.~\ref{#1}}
\def\App#1{App.~\ref{#1}}
\graphicspath{
	{./figures/}
	{../figures/}
}

\usepackage{xifthen}
\usepackage{xcolor}

\newcommand{\gettitle}{Dissecting the moat regime at low energies I:\\ Renormalization and the phase structure}

\hypersetup{
	colorlinks,
	linkcolor={blue!75!black},
	citecolor={blue!75!black},
	urlcolor={blue!75!black}, 
	pdftitle={\gettitle},
	pdfauthor={},
	pdfkeywords={}
	{} 
	bookmarksopen=true,
	bookmarksopenlevel=2,
	bookmarksnumbered=true
}

\begin{document}
\title{\gettitle}

\author{Fabian Rennecke \,\orcidlink{0000-0003-1448-677X}}
\email{fabian.rennecke@theo.physik.uni-giessen.de}
\affiliation{Institute for Theoretical Physics, Justus Liebig University Giessen, 35392 Giessen, Germany}
\affiliation{Helmholtz Research Academy Hesse for FAIR (HFHF), Campus Giessen, Giessen, Germany}

\author{Shi Yin \,\orcidlink{0000-0001-5279-6926}}
\email{shi.yin@theo.physik.uni-giessen.de}
\affiliation{Institute for Theoretical Physics, Justus Liebig University Giessen, 35392 Giessen, Germany}

\begin{abstract}

Dense QCD matter can feature a moat regime, where the static energy of mesons is minimal at nonzero momentum. Valuable insights into this regime can be gained using low-energy models. This, however, requires a careful assessment of model artifacts. We therefore study the effects of renormalization and in-medium modifications of quark-meson interaction on the moat regime. To capture the main effects, we use a two-flavor quark-meson model at finite temperature and baryon density in the random phase approximation. We put forward a convenient renormalization scheme to account for the nontrivial momentum dependence of meson self-energies and discuss the role of renormalization conditions for renormalization group consistent results on the moat regime. In addition, we demonstrate and that its extent in the phase diagram critically depends on the interaction of quarks and mesons.

\end{abstract}

\maketitle
	
\section{Introduction}

In recent decades, research into quantum chromodynamics (QCD) phase transitions has deepened progressively. Study on the QCD phase structure currently has primarily focused on the search for the QCD critical endpoint (CEP). The completion of phase I and II of the Beam Energy Scan (BES) at the Relativistic Heavy Ion Collider (RHIC) is expected to provide further insights into the CEP \cite{Chen:2024aom,Mondal:2024sil}. Given that no compelling signals of the CEP have been observed within the collision energy range of 7.7 to 200 GeV \footnote{STAR has recently reported a significant deviation of the kurtosis of the net-proton distribution from the noncritical baseline at $\sqrt{s} = 19.6$\,GeV \cite{STAR:2025zdq}. This corresponds to $(T,\mu_B) \approx (156,200)$\,MeV at freeze-out and is therefore well within the range where a CEP has been excluded both from functional and lattice QCD, see, e.g., Refs.\ \cite{Fu:2019hdw, Gao:2020fbl, Gunkel:2021oya, Bazavov:2017dus, Borsanyi:2025dyp}.}, heavy-ion experiments are gradually shifting toward lower-energy collisions, e.g., fixed-target collisions \cite{STAR:2022etb,CBM:2016kpk,Mackowiak-Pawlowska:2016qon,Kapishin:2020cwk,Sako:2019hzh}, which will also offer valuable information on the phase structure at large density. In addition to its fundamental and astrophysical significance, this shift provides additional motivation for theoretical explorations of the rich phase structure at higher densities.

In the low-density region, lattice QCD simulations have revealed that the chiral phase transition is a smooth crossover \cite{Bellwied:2015rza,Bazavov:2018mes}. Functional continuum approaches to QCD based on the functional renormalization group (FRG) \cite{Fu:2019hdw} and Dyson-Schwinger Equations (DSE) \cite{Gao:2020fbl, Gunkel:2021oya} have provided strikingly consistent predictions for the location of the CEP. Additionally, as the chemical potential increases, a moat regime, introduced in Ref.\ \cite{Pisarski:2021qof}, where the static (zero frequency) dispersion of mesons develops a minimum at nonzero momentum, has been found using the FRG \cite{Fu:2019hdw, Fu:2024rto}. 
This suggests the possibility of a richer phase structure at intermediate to high density, and provides motivation for the ongoing searches for inhomogeneous phases \cite{Motta:2023pks, Motta:2024agi, Motta:2024rvk, Fu:2024rto}. In fact, the moat regime can be identified as a common feature of various systems with inhomogeneous phases and oscillatory regimes, e.g., \cite{Fulde:1964zz, Hornreich:1975zz, Seul-Andelman:1995, Chakrabarty_2011, Sedrakyan:2013qja, Buballa:2014tba, Schindler:2019ugo, Pisarski:2020dnx, Schindler:2021otf, Nussinov:2024erh}. Furthermore, it may lead to observable signals in heavy-ion collisions \cite{Pisarski:2020gkx, Pisarski:2021qof, Rennecke:2021ovl, Rennecke:2023xhc, Nussinov:2024erh}.

Given its general nature, it is worthwhile to explore the moat regime not only directly in QCD \cite{Fu:2019hdw, Fu:2024rto}, but also in effective models. On the one hand, because some of its features may be easier to understand in simpler models. On the other hand, because the resulting insights could also be valuable for other systems. Thus far, the moat regime has been studied in Nambu--Jona-Lasinio (NJL) models \cite{Koenigstein:2021llr, Pannullo:2023one, Pannullo:2023cat, Koenigstein:2023yzv, Winstel:2024dqu, Pannullo:2024sov} and Quark-Meson (QM) models \cite{Topfel:2024iop, Cao:2025zvh} for quark matter, and the Quark-Meson Coupling (QMC) model \cite{Motta:2025xop} for nuclear matter. All these model studies have been performed using some form of random phase approximation (RPA), i.e., solving the gap equation for the chiral condensate neglecting bosonic fluctuations (mean-field) and computing the boson self-energy arising from fermions at the one-loop level with the fermion masses that follow from the gap equation. Curiously, the moat regime seems to exist at large temperature even at vanishing density in the QM model in RPA \cite{Topfel:2024iop, Cao:2025zvh}. One shortcoming of such an approximation is that in-medium modifications of quark-meson Yukawa interactions are neglected. We will show that these modifications turn out to be important for the moat regime at finite temperature and density. 

Since the relevant low-energy degrees of freedom in such models are not emergent like in QCD, but are put in at the outset, there is an inherent ultraviolet cutoff scale and renormalization plays an important role. If not treated properly, large renormalization scheme and scale dependencies may contaminate the results. This has been pointed out, e.g., in NJL models in Refs.\ \cite{Partyka:2008sv, Buballa:2020nsi, Pannullo:2023cat, Pannullo:2024sov}. Clearly, only renormalization scale independent results, also called RG-consistent in the present context \cite{Braun:2018svj}, have a chance to be reliable. 

In the present work, we address the issue of renormalization and in-medium modifications of interactions in the phase diagram of a two-flavor QM model in $(3\!+\!1)$ spacetime dimensions, which is known to support a moat regime \cite{Topfel:2024iop}. QM models have been studied in great detail using various methods, e.g., in Refs.\ \cite{Gell-Mann:1960mvl, Jungnickel:1995fp, Schaefer:2004en, Tripolt:2013jra, Pawlowski:2014zaa, Kovacs:2016juc}, as they can provide valuable information in particular regarding chiral physics of QCD. In addition, it has been shown that these models naturally arise as low-energy models of QCD \cite{Braun:2014ata, Rennecke:2015eba, Rennecke:2015lur, Cyrol:2017ewj, Fu:2019hdw, Ihssen:2024miv}.

Since the moat regime is reflected in the momentum dependence of meson self-energies, renormalization of this part needs to be done with care. The spatial wave function renormalization, which is negative in the moat regime, is power-counting marginal and hence needs to be renormalized. It is therefore clear that an additional renormalization condition is required in order to achieve RG consistency. To this end, we set up an efficient renormalization scheme that facilitates on-shell renormalization and, in addition, mends an unphysical large-momentum behavior of the meson self-energies in RPA. Any renormalization scale dependence of the location of the moat regime in the phase diagram is fully removed this way. This is similar to the case of the phase diagram at finite isospin density \cite{Brandt:2025tkg} and in the presence of diquarks \cite{Gholami:2025afm}. And on-shell renormalization of the QM model has also been discussed in Refs.\ \cite{Carignano:2014jla, Carignano:2016jnw, Adhikari:2016eef, Adhikari:2017ydi, Buballa:2020xaa, Rai:2022wth}.
Furthermore, we show that the appearance of the moat regime at vanishing density found in Refs.\ \cite{Topfel:2024iop, Cao:2025zvh} is an approximation artifact which is alleviated once in-medium modifications are taken into account. We clarify that the competition between creation-annihilation and particle-hole processes in the hot and dense medium is ultimately responsible for these observations.

This paper is organized as follows:  In \Sec{sec:LEFT}, we present the setup of the QM model in RPA. In \Sec{sec:regularization}, we discuss the regularization and renormalization, with special attention to the momentum-dependent corrections relevant for the moat regime. This is applied to the phase diagram at finite temperature and quark chemical potential in \Sec{sec:dissecting}. There, we clarify the large temperature behavior of the spatial pion wave function renormalization and the underlying microscopic effects, demonstrate the importance of proper renormalization for the phase diagram and study the effect of in-medium modifications of the quark-meson Yukawa interaction. \Sec{sec:summary} is devoted to the conclusions and the appendices to technical details.

\section{Setup}\label{sec:LEFT}
\subsection{Quark-meson model}\label{sec:qm_model}
We employ a QM model in the mean-field approximation to investigate the moat regime at finite temperature and density. As we shall see, this allows us to dissect the moat regime in a clear and relatively simple fashion. The renormalized Lagrangian of the QM model with two light quark flavors and (pseudo-) scalar bound states, $\pi$ and $\sigma$, is in Euclidean space given by
%
\begin{align}\label{eq:L}
\begin{split}
\mathcal{L}[\phi,q,\bar{q}]&=\bar{q}\big[\gamma_\mu\partial_\mu-\gamma_0 \hat{\mu}\big]q+\frac{1}{2}\,(\partial_\mu\phi)^2\\[2ex]
&\quad+h\,\bar{q}(T^0\sigma+i\gamma_5\mathbf{T}\cdot\boldsymbol{\pi})q+U(\rho)-c\sigma\\[2ex]
&\quad+ \mathcal{L}_{\rm ct}\,.
\end{split}
\end{align}
%
The $q$ and $\bar{q}$ denote the quark and anti-quark Dirac spinors respectively. $\hat{\mu}=\mathrm{diag}(\mu_u,\mu_d)$ is the chemical potential of the light quarks. In this work we ignore the difference between $u$ and $d$ quarks, so their chemical potentials are set to be same here $\mu_u=\mu_d\equiv\mu$. The meson field is defined by $\phi=(\sigma,\boldsymbol{\pi})$ and $\rho=\phi^2/2$. $h$ is the Yukawa coupling which determines the interaction strength between fermions and bosons. Here we ignore the difference between the pion-quark and sigma-quark interactions and use the pion channel for all computations. $T^0=\frac{1}{\sqrt{2N_f}}\mathbbm{1}_{N_f\times N_f}$ and $\mathbf{T}=\frac{1}{2}\boldsymbol{\tau}$ are $SU(N_f)$ flavor space generators with $N_f=2$ here, so $\boldsymbol{\tau}$ are the Pauli matrices.  The bare mesonic potential is given by $U(\rho)$, with a linear symmetry breaking term $-c\sigma$, reflecting the nonzero current quark masses. $\mathcal{L}_{\rm ct}$ contains the counter terms, which are specified in \Sec{sec:regularization}.

In the mean-field approximation, quarks are integrated out while meson fluctuations are neglected. Hence, the bare meson potential $U(\rho)$ in \Eq{eq:L} receives quantum corrections only from the functional determinant of quarks. This leads to the full effective potential
\begin{align}\label{eq:vdet}
V(\rho) = U(\rho) - \frac{T}{\mathcal{V}} \ln\det \mathcal{M}(\sigma)\,,
\end{align}
with the spatial volume $\mathcal{V}$ and the Dirac operator
\begin{align}
\mathcal{M}(\sigma) = \gamma_\mu\partial_\mu - \gamma_0\mu + h T^0 \sigma\,.
\end{align}
We used that homogeneous chiral symmetry breaking of isospin symmetric matter implies that the mesonic mean field is scalar, so $\rho = \sigma^2/2$. As we will see below, inhomogeneous chiral symmetry breaking will not play a role here. $V(\rho)$ can be split into two contributions,
%
\begin{align}
V(\rho)=V_{\mathrm{vac}}(\rho)+V_{\mathrm{thermal}}(\rho)\,.\label{eq:ep}
\end{align}
%
$V_{\mathrm{vac}}$ is the vacuum contribution and $V_{\mathrm{thermal}}$ gives the in-medium corrections from finite temperature and chemical potential. 

The vacuum part is divergent and requires renormalization. This will be discussed in \Sec{sec:regularization}. The thermal part of the potential reads
%
\begin{align}
&V_{\mathrm{thermal}}(\rho)=\frac{N_cN_f}{3\pi^2}\int_0^\infty\! dq \,\frac{q^4}{E_q}\nonumber\\[2ex]
&\times\bigg[\big(n_F(E_q;T,\mu)+n_F(E_q;T,-\mu)\big)+\nu\rho-\frac{\lambda}{2}\rho^2\bigg]\,,\label{eq:ep_thermal}
\end{align}
%
where we used the ansatz $U(\rho) = -\nu\rho+\lambda\rho^2/2$ for the mesonic part of the potential. The parameters will be given in \Sec{sec:regularization}. $N_c=3$ is the number of colors.  $q$ stands for the magnitude of the internal spatial momentum. The summation over the Matsubara frequencies of the imaginary time formalism, and the angular integration have already been carried out in this expression. $m_f^2=h^2\rho/2$ is the squared constituent quark mass. The energy of the quark is given by $E_q=\sqrt{q^2+m_f^2}$. $n_F(x;T,\pm\mu)=1/(\mathrm{exp}((x\mp\mu)/T)+1)$ is the Fermi-Dirac distribution of (anti-) quarks. While chiral symmetry is broken with our ansatz for $U(\rho)$ in the vacuum, the thermal corrections induced by these distributions will restore the symmetry at sufficiently high temperature.

The pion and sigma curvature masses are obtained from derivatives of the effective potential with respect to $\rho$, $m^2_\pi=V'(\rho)$ and $m^2_\sigma=V'(\rho)+2\rho V''(\rho)$. All physical observables are defined for field values $\rho = \rho_0$, where $\rho_0$ is the solution of the gap equation
%
\begin{align}
\frac{\partial}{\partial\rho}\Big[V(\rho)-c(2\rho)^\frac{1}{2}\Big]\bigg|_{\rho=\rho_0}=0\,.\label{eq:eom}
\end{align}
%
We emphasize that this low-energy effective model cannot give us quantitatively reliable results for QCD. However, we will demonstrate that it can still give us valuable qualitative insights into the moat regime at finite temperature and density.

Now that we have a complete setup for the QM model, we will discuss the main probes of the moat regime: the two-point function and wave function renormalization of pions.

\subsection{Pion two-point function}\label{sec:pion_prop}

\begin{figure}[t]
\includegraphics[width=0.49\textwidth]{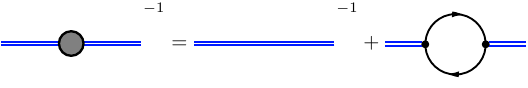}
\caption{The full meson propagator, denoted by the gray dot, as computed in this work. The blue double-lines and the black lines stand for free meson and quark propagators, respectively.}\label{fig:feynman-dia-meson}
\end{figure}

From previous studies , e.g., \cite{Fu:2019hdw,Pisarski:2021qof,Koenigstein:2021llr, Rennecke:2023xhc, Pannullo:2024sov, Fu:2024rto, Topfel:2024iop, Cao:2025zvh, Motta:2025xop}, we know that the moat regime manifests itself in the non-monotonic spatial momentum dependence of two-point correlation functions of bosonic states. This is signaled by a negative value of the meson wave function renormalization at vanishing momentum. Our starting point is therefore the pion two-point function,
%
\begin{align}\label{eq:sigma_pion}
\Sigma_\pi(p^2;T,\mu)=p^2+m^2_\pi+\Pi^\pi_{\mathrm{RPA}}(p^2;T,\mu)\,,
\end{align}
%
with the four-momentum $p=(p_0,\boldsymbol{p})$.
The first two terms on the right-hand side are the free pion two-point function and the last term is the one-loop self-energy correction. Here the free pion mass is given by $m^2_\pi=\lambda\rho-\nu$. The diagrammatic representation of \Eq{eq:sigma_pion} is shown in \Fig{fig:feynman-dia-meson}. The first term on the right-hand side is the bare inverse pion propagator and the second term shows the self-energy, where we only take the quark loop contribution into account. Since we evaluate this diagram on the solution of the gap equation, this corresponds to a random phase approximation (RPA).

%
%

It is convenient to decompose the two-point function into momentum-dependent and independent parts. At nonzero temperature and density, we may use the parametrization
%
\begin{align}\label{eq:sipi}
\Sigma_\pi(p^2;T,\mu)=&Z^\|_\pi(p^2;T,\mu)\,p_0^2+Z^\perp_\pi(p^2;T,\mu)\,\boldsymbol{p}^2\nonumber\\[2ex]
                      &+\bar m^2_\pi(T,\mu)\,.
\end{align}
%
$Z^{\parallel,\perp}_\pi(p^2;T,\mu)$ are momentum-dependent wave function renormalizations in temporal and spatial direction, and $\bar m_\pi^2 = m_\pi^2 + \Pi^\pi_{\mathrm{RPA}}(0;T,\mu)$ is the dressed pion curvature mass. The difference between the parallel and transverse components are caused by the spacetime $O(4)$ symmetry breaking at finite temperature; for more discussion within the FRG approach, see, e.g., \cite{Yin:2019ebz}. We are primarily interested in static properties here. Of central interest is the spatial/transverse wave function renormalization of pions,
%
\begin{align}
Z^\perp_\pi(p=0;T,\mu)=&\frac{\partial}{\partial\boldsymbol{p}^2}\Sigma_\pi(p_0=0,\boldsymbol{p};T,\mu)\bigg|_{\boldsymbol{p}^2=0}\,,
\label{eq:Zprojection}
\end{align}
%
as a negative $Z^\perp$ signals the moat regime and pions, being the lightest hadrons, are its most sensitive probe \cite{Pisarski:2021qof}.
The spatial components of the wave function renormalization reflect the space-like properties of the mesonic two-point function, and this is where the moat behavior occurs \cite{Fu:2024rto}.

With these definitions, we can give the equation for the pion two-point function and the pion wave function renormalization at finite temperature and quark chemical potential. The self-energy in \Eq{eq:sigma_pion} and \Fig{fig:feynman-dia-meson} in the QM model is given by
%
\begin{align}
&\Pi^\pi_{\mathrm{RPA}}(p^2;T,\mu)\nonumber\\[2ex]
=&-h^2N_c\int\frac{d^3q}{(2\pi)^3}\bigg[2\mathcal{F}_{(1)}(q)-p^2\mathcal{FF}_{(1,1)}^-(p,q)\bigg]
\,.\label{eq:pion-two-point}
\end{align}
%
The first term in the square bracket is the momentum independent part of the two-point function, which is related to the correction of the bare pion curvature mass. The second, momentum dependent term gives rise to a nontrivial pion wave function renormalization. $\mathcal{F}_{(1)}$ and $\mathcal{FF}_{(1,1)}^-$ are the threshold functions of the quark loop, and are given in \App{app:thre_fun}.

Combining the projection in \Eq{eq:Zprojection} with \Eq{eq:pion-two-point}, we obtain the transverse pion wave function renormalization at vanishing external momentum,
%
\begin{align}
\begin{split}
&Z^\perp_\pi(p=0;T,\mu)\\[2ex]
&\quad =1-\frac{h^2N_c}{\pi^2}\int dq\,q^2\Big[-\mathcal{F}_{(2)}(q^2)+\frac{2}{3}q^2\mathcal{F}_{(3)}(q^2)\Big]
\,.\label{eq:Zpi}
\end{split}
\end{align}
%
The threshold functions here are also given in \App{app:thre_fun}. Similar to the equation of the effective potential, \Eq{eq:ep}, the equation of both, the two-point function and the wave function renormalization can also be divided into vacuum part and thermal parts,
%
\begin{align}
Z^\perp_\pi=Z^\perp_{\pi,\mathrm{vac}}+Z^\perp_{\pi,\mathrm{thermal}}\,,
\label{eq:Z_v_t}
\end{align}
%
and
%
\begin{align}
\Pi^\pi_\mathrm{RPA}=\Pi^\pi_\mathrm{vac}+\Pi^\pi_\mathrm{thermal}\,.
\label{eq:sigma_v_t}
\end{align}
%
Both vacuum parts are also divergent and need to be regularized. The regularization and renormalization procedures will be introduced in next subsection.

The self-energy correction in \Eq{eq:pion-two-point} and the wave function renormalization in \Eq{eq:Zpi} depend on the Yukawa interaction $h$. In RPA it is not renormalized, however, its behavior turns out to be crucial for the size of the moat regime, which will be discussed in \ref{subsec:yukawa}.

\section{Regularization and renormalization}\label{sec:regularization}

As mentioned above, the vacuum contributions in both the effective potential and the two-point function require regularization. We choose the dimensional regularization here. This allows us to analytically separate the divergent part of the integral, while the finite in-medium part can be fully included. Of course, the results may depend on the choice of renormalization scale. We will discuss this later in this section.

First, we consider the vacuum part of effective potential \Eq{eq:ep} and separate the divergent part of the integral from the convergent part. From the quark determinant \eq{eq:vdet} we obtain the well-known result in $3-2\epsilon$ dimensions, e.g., \cite{Skokov:2010sf, Mukherjee:2021tyg},
%
\begin{align}
\begin{split}
V_{\mathrm{vac}}(\rho) &= \frac{N_f N_c m_f^4}{16 \pi^2}\bigg[\frac{1}{\epsilon}-2 \ln\!\Big(\frac{m_f}{M}\Big)+C + \mathcal{O}(\epsilon)\bigg]\\[2ex]
&\quad + \mathcal{L}_{\rm ct}\big|_{\phi = {\rm const.}}\,,
\end{split}
\end{align}
%
with $C = \ln 4\pi-\gamma_E + \frac{3}{2}$, where $\gamma_E$ is Euler's constant. As detailed below, after proper renormalization, our results will not depend on the renormalization scale $M$ \cite{Skokov:2010sf}.

We start by discussing the modified minimal subtraction ($\overline{\rm MS}$) renormalization scheme. We hence include the counter term for the quartic meson coupling
\begin{align}\label{eq:ctV}
    \mathcal{L}_{\rm ct} \supset \delta_\lambda \phi^4\,,\qquad \delta_\lambda = -\frac{N_c N_f h^4}{2^8 \pi^2} \bigg(\frac{1}{\epsilon} + C \bigg)\,,
\end{align}
which leads to
%
\begin{align}
V^{\overline{{\rm MS}}}_{\mathrm{vac}}(\rho)=-\frac{N_cN_fm_f^4}{8\pi^2}\,\ln\!\Big(\frac{m_f}{M}\Big)\,.
\label{eq:ep_re}
\end{align}
%
For different values of the renormalization scale $M$, we can adjust the values of parameters $\nu$ and $\lambda$ in \Eq{eq:ep_thermal} to ensure that the physical values, e.g., the meson masses, the constituent quark mass and the chiral condensate $\sigma_0$, in the vacuum remain unchanged. With the renormalization conditions specified below, the chiral phase transition turns from a crossover to the first order phase transition at a CEP around $\mu_{\rm CEP}=290\,\mathrm{MeV}$ and $T_{\rm CEP} = 30\,\mathrm{MeV}$. We explicitly checked that the position of the CEP stays the same at renormalization scale equal to 300, 400 and 500 MeV if we fix the meson mass $\bar m_{\pi/\sigma}$, quark mass $m_f$ and chiral condensate $\sigma_0$ at vacuum. This also confirms the previous statement that thermodynamics does not depend on the choice of renormalization scale. For $M=300$ MeV, we use the following parameters: $\nu=(482\,\mathrm{MeV})^2$, $\lambda=75.8$, $c=0.0017\,\mathrm{GeV}^3$ and $h=6.5$. These parameters correspond to $\sigma_0 = f_\pi =92$ MeV, $\bar m_\pi^{\rm vac}=136$ MeV, $\bar m_\sigma^{\rm vac}=480$ MeV and $m_f^{\rm vac}=300$ MeV as renormalization conditions in vacuum. 

These conditions are incomplete without discussing the regularization of the two-point functions and the wave function renormalizations. Using \Eq{eq:pion-two-point}, performing dimensional regularization and taking into account the counter-term, the vacuum part of the pion self-energy is
%
\begin{align}
&\Pi_{\mathrm{vac}}^{\pi}(p^2;m_f)= \frac{\partial^2 \mathcal{L}_{\rm ct}}{\partial \phi^2} \nonumber\\[2ex]
&-\frac{h^2N_c}{8\pi^2}\Bigg\{m_f^2\bigg[-\frac{1}{\epsilon}-1+\gamma_E-\ln 4\pi+2\ln\!\Big(\frac{m_f}{M}\Big)\bigg]\nonumber\\[2ex]
&\qquad\quad\quad-\frac{1}{2}p^2\bigg[\frac{1}{\epsilon}-\gamma_E+\ln 4\pi+2-2\ln\Big(\frac{m_f}{M}\Big)\nonumber\\[2ex]
&\qquad\quad\quad-\frac{\sqrt{p^2+4m^2_f}}{p}\ln\!\bigg(\frac{\sqrt{p^2+4m^2_f}+p}{\sqrt{p^2+4m^2_f}-p}\bigg)\bigg]\Bigg\}\,.\label{eq:momentum_1loop}
\end{align}
%
The formula consists of two parts, a momentum-independent part (the second row) related to the rest mass of the pion, and a momentum-dependent part (the third row) related to a momentum-dependent wave function renormalization. It is clear that renormalization of the latter part requires an additional, kinetic counterterm, 
\begin{align}\label{eq:deltaZ}
    \mathcal{L}_{\rm ct} \supset \frac{1}{2}\delta_Z \big(\partial_\mu\phi\big)^2\,.
\end{align}
Note that in a medium this may split into a temporal and a spatial part, $\delta_Z\, p^2 \rightarrow \delta_Z^\parallel\, p_0^2+\delta_Z^\perp\, \boldsymbol{p}^2$, because of Lorentz symmetry breaking.

From the projection introduced in \Eq{eq:Zprojection} we can extract the regularized equation for pion wave function renormalization at vanishing momentum,
\begin{align}\label{eq:Zpiren}
Z^\perp_{\pi,\mathrm{vac}}(0) &=
\frac{h^2N_c}{16\pi^2}\Bigg[ \frac{1}{\epsilon} - \gamma_E + \ln 4\pi - 2 \ln\bigg(\frac{m_f}{M}\bigg) \Bigg]\nonumber\\[2ex]
&\quad +1+\delta_Z
\end{align}

%
\begin{figure}[t]
\includegraphics[width=0.49\textwidth]{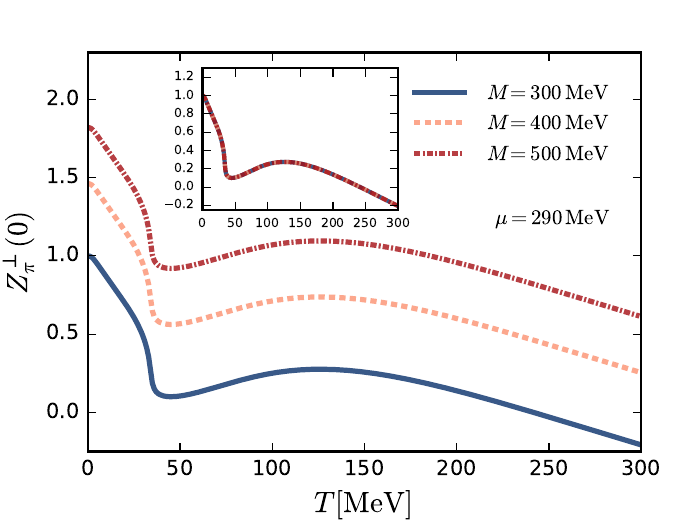}
\caption{Spatial pion wave function renormalization at vanishing momentum as function of temperature for different renormalization scales $M=300$, $400$ and $500$ MeV for $\bar C = 0$ in \Eq{eq:Zpi_re}. The inset gives the renormalized result using the condition in \Eq{eq:Zcond}.}\label{fig:ZpiM}
\end{figure}
%

The momentum-independent part of the self-energy is renormalized by the same counter-term as the effective potential, \Eq{eq:ctV}.
For the kinetic counter-term we choose
\begin{align}\label{eq:delZ}
    \delta_Z = -\frac{h^2 N_c}{16\pi^2} \bigg[ \frac{1}{\epsilon} - \gamma_E + \ln 4\pi - 2 \bar C \bigg]\,,
\end{align}
where $\bar C = \bar C(M)$ is a renormalization scale dependent constant that we use to adjust to the renormalization condition specified below. This leads to the renormalized self-energy and spatial wave function renormalization,
%
\begin{align}\label{eq:two_point_re}
\Pi_\mathrm{vac}^{\pi,\mathrm{re}}(p^2;m_f)=&-\frac{h^2N_c}{8\pi^2}\Bigg\{m_f^2\bigg[\frac{1}{2}+2\ln\!\Big(\frac{m_f}{M}\Big)\bigg]\nonumber\\[2ex]
&+p^2\Bigg[\bar C -1 + \ln\!\Big(\frac{m_f}{M}\Big)\nonumber\\[2ex]
&+\frac{\sqrt{p^2+4m_f^2}}{2p}\ln\!\Bigg(\frac{\sqrt{p^2+4m_f^2}+p}{\sqrt{p^2+4m_f^2}-p}\Bigg)\Bigg]\Bigg\}\,,
\end{align}
%
and
%
\begin{align}\label{eq:Zpi_re}
Z^{\perp,\mathrm{re}}_{\pi,\mathrm{vac}}(0)= 1-\frac{h^2N_c}{8\pi^2}\bigg[\bar C + \ln\!\Big(\frac{m_f}{M}\Big)\bigg]\,.
\end{align}
To ensure independence of our results on the renormalization scale $M$, we need to identify suitable renormalization conditions. 
For the effective potential we use physical quantities, like meson masses and decay constants, to fix the parameters for any $M$, i.e., their RG running. We can directly use the $\overline{\rm MS}$-result for the effective potential in \Eq{eq:ep_re} and adjust our model parameters to reproduce the physical quantities as specified above. While this procedure is common in the literature, it is imprecise without further conditions, because the curvature masses are identified with the measured masses. A more precise procedure is to do a proper on-shell renormalization, and to identify the pole masses $m_{\rm p}$, defined via $\Sigma(p_0=i m_p, \boldsymbol{p} = 0;T,\mu) = 0$, with the physical masses \cite{Carignano:2014jla, Adhikari:2016eef}. In a low-frequency expansion, this fixes the ratio $\bar m_\pi^2/Z^\parallel_\pi$, where $Z^\parallel_\pi = Z^\parallel_\pi(p = 0)$.

%
\begin{figure*}[t]
\includegraphics[width=0.49\textwidth]{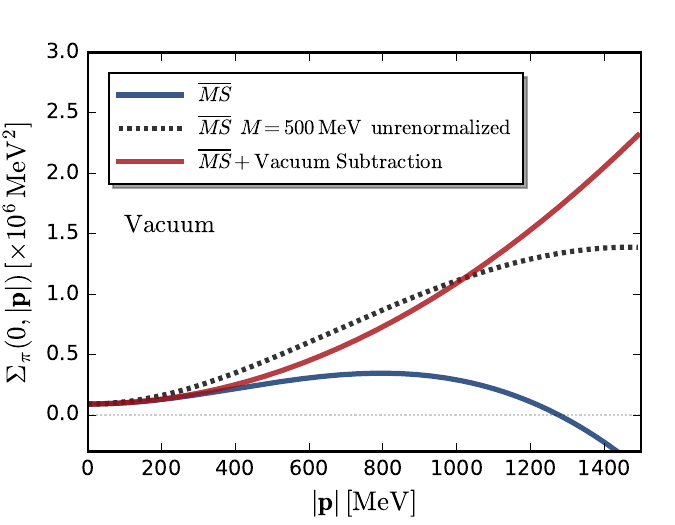}
\includegraphics[width=0.49\textwidth]{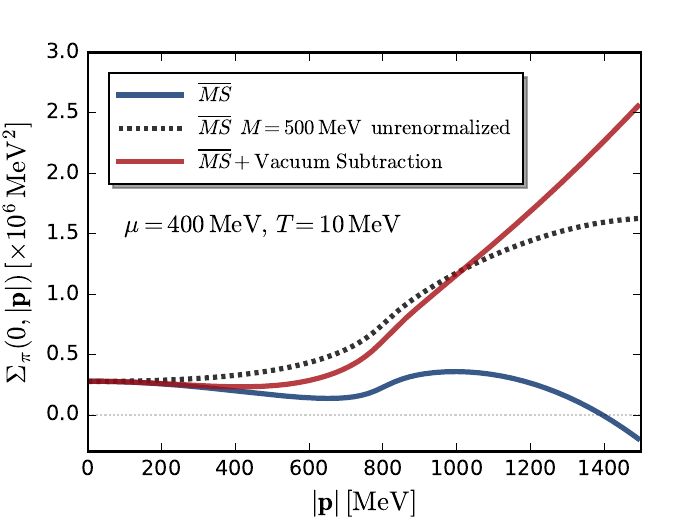}
\caption{Pion self-energy as function of the spatial momentum in vacuum (left) and at $\mu$=400 MeV, T=10 MeV (right) using the conventional $\overline{\rm MS}$ scheme (blue solid line) and the scheme with additional vacuum subtraction defined in \Eq{eq:two_point_re2} (dark red solid line). In both cases, the renormalization condition in \Eq{eq:Zcond} is used, rendering the self-energy independent of the renormalization scale $M$. The black dotted line shows the result without renormalization at $M=500$\,MeV.}\label{fig:renorm_compare}
\end{figure*}
%

Since we need an additional counter term for the momentum-dependent part of the self-energy,  \Eq{eq:deltaZ}, a further renormalization condition is required. Noting that, based on \Eq{eq:sipi}, the meson propagator in a low-momentum expansion can in general be written as
\begin{align}\label{eq:Gphi}
    G_\phi = \frac{1/Z^\parallel}{p_0^2+\frac{Z^\perp}{Z^\parallel}\boldsymbol{p}^2 + \frac{\bar m^2_\phi}{Z^\parallel} + \cdots}\,,
\end{align}
where the dots denote higher-momentum corrections, and that $Z^\perp = Z^\parallel$ in vacuum, $Z_\perp$ determines the residue of the propagator at the mass pole in vacuum. One can hence use standard on-shell renormalization and use the residue as an additional renormalization condition \cite{Peskin:1995ev}, see also Refs.\ \cite{Adhikari:2016eef, Brandt:2025tkg} and \footnote{While there is no explicit renormalization condition for the wave function renormalization in Refs.\ \cite{Carignano:2014jla, Carignano:2016jnw, Buballa:2020xaa}, $Z_\pi$ is implicitly fixed through $f_\pi$. The advantage of an explicit condition for the wave function renormalization is that it straightforwardly generalizes also to other mesons.}. $Z^\perp/Z^\parallel$ can be interpreted as the squared group velocity of the meson. Owing to Lorentz invariance, it is always one in vacuum, regardless of the renormalization condition.

The screening mass $m_s$, defined by $\Sigma(p_0=0, \boldsymbol{p} = im_s;T,\mu) = 0$, is given by $\bar m_\pi^2/Z_\pi^\perp$ in the low-momentum expansion. One could therefore also use a combination of measured vacuum pole masses and finite-temperature screening masses from lattice QCD, e.g., from Ref.\ \cite{Bazavov:2019www}, to fix $Z_\pi^\parallel$ and $Z_\pi^\perp$ \footnote{Note that $Z_\pi^\parallel = Z_\pi^\perp$ in vacuum, and this remains a good approximation even at moderate $T$ and $\mu$ \cite{Yin:2019ebz}. However, it clearly breaks down when the system enters the moat regime, as $Z^\perp$ changes sign while $Z^\parallel$ is always positive due to causality.}.

Here, go into the former direction and renormalize the spatial wave function renormalization to one in vacuum,
\begin{align}\label{eq:Zcond}
    Z_\pi^\perp(p=0;T=0,\mu=0) = 1\,.
\end{align}
This condition fixes the parameter
\begin{align}
    \bar C(M) = \ln(M/m_f^{\rm vac})
\end{align}
in \Eq{eq:delZ}. From \Eq{eq:Gphi} follows that in case of only a mild momentum dependence of the self-energy, this condition enforces curvature, pole and screening masses to be identical in vacuum. We will generalize this to an arbitrary momentum dependence below, but note here that \Eq{eq:Zcond} is clearly not the conventional on-shell renormalization condition for the residue. The residue is fixed, by definition, at the pole, so the condition $\partial_{p^2} \Pi^\pi_{\rm vac}(p^2)\big|_{p = i m_{p,\pi}} = 0$ (with the appropriate analytic continuation) fixes it to one. Our condition in \Eq{eq:Zcond} is geared towards the moat regime, as this is defined by $Z^\perp$ at vanishing momentum.

In any case, proper renormalization leads to RG-consistency, i.e., independence of our results on $M$. This is illustrated in \Fig{fig:ZpiM}, where we show $Z_\pi^\perp$ for different renormalization scales. Simply setting $\bar C  = 0$, the results show a strong $M$ dependence. In contrast, by taking into account the running of $Z_\pi^\perp$ by adjusting $\bar C(M)$ to enforce \Eq{eq:Zcond} for different $M$, our results are clearly RG consistent. This is shown in the inset of \Fig{fig:ZpiM}. Note that all the lines collapse onto the result with $M=300$\,MeV because this agrees with the vacuum quark mass, so the renormalization scale dependent part of \Eq{eq:Zpi_re} vanishes trivially. Our analysis implies that at least some of the scheme dependencies discussed in the recent literature are likely due to incomplete renormalization of the self-energy, see, e.g., \cite{Koenigstein:2023yzv, Pannullo:2024sov}.

It may seem that renormalization of the effective potential and the momentum-dependent part of the self-energy work differently here. On the one hand, for the effective potential we minimally subtract the divergent contribution and then adjust the model parameters according to our renormalization conditions. This naturally gives rise to running coupling, i.e., $\nu = \nu(M)$ and $\lambda = \lambda(M)$. On the other hand, for the momentum-dependent part we have an $M$-dependent counter term. Note, however, that we write the QM model Lagrangian in \Eq{eq:L} with a trivial pion wave function renormalization $Z_\pi = 1$. But since a nontrivial $Z_\pi$ is generated in RPA, we might as well introduce it already in the effective Lagrangian \Eq{eq:L}. Any renormalization condition on $Z_\pi$ would then naturally also lead to a running, $Z_\pi = Z_\pi(M)$. Hence, renormalization, as expected, always works the same.

Since the moat regime is entered for $Z^\perp\leq 0$, this demonstrates the importance of proper renormalization. The thermal contributions to $Z^\perp$, which are given in Eqs.\ \eq{eq:threshold_fun_f2}, \eq{eq:threshold_fun_f3} and \eq{eq:threshold_fun_p}, are independent of $M$ and always negative. Hence, where the system enters the moat regime crucially depends on the vacuum part and, therefore, on the renormalization condition. We emphasize that this ambiguity is inherent only to low-energy models. In QCD, all meson correlations, including their self-energies, are uniquely defined, emergent quantities that are fully determined by microscopic quark and gluon interactions \cite{Braun:2014ata, Rennecke:2015eba, Fu:2019hdw, Fu:2024rto}.

At sufficiently large momentum $p$ the $\overline{\rm MS}$ vacuum contribution to the self-energy in \Eq{eq:two_point_re} also turns negative, and becomes increasingly negative for even larger $p$. Hence, the two-point function \eq{eq:sigma_pion} is bound to become negative at some large $p$, signaling an instability of the system at any $T$ and $\mu$. This is demonstrated by the blue solid lines in \Fig{fig:renorm_compare}. In the right panel, the nonmonotonic momentum dependence of $\Sigma_\pi$ for $p\lesssim 900\, {\rm MeV}$ signals the moat regime. The behavior at larger $p$ is the aforementioned instability. Note that this is manifestly different from an instability towards an inhomogeneous phase, where the two-point function is zero or negative at the bottom of the moat \cite{Fu:2024rto}. From the left panel, one can see that even in vacuum, the two-point function exhibits this instability at large momentum, indicating that it is unphysical.

The solid lines in \Fig{fig:renorm_compare} show the renormalized self-energy using the condition in \Eq{eq:Zcond} both in the vacuum (left) and in the moat regime (right). Note that this is sufficient to render the full self-energy independent of the renormalization scale, as all $M$-dependencies in \Eq{eq:two_point_re} are removed by the renormalization conditions for $\nu$, $\lambda$ and $Z_\pi^\perp$. In contrast, without the renormalization condition in \Eq{eq:Zcond} a strong renormalization scale dependence remains, as exemplified by the black dotted line in \Fig{fig:renorm_compare}.

It is perhaps not too surprising that the results of our calculation are only reliable for a limited range of momenta in the presence of a finite renormalization scale $M$. This is similar to the large logarithms encountered in perturbation theory. To remedy this to some extend, we introduce a new renormalization procedure where we supplement the $\overline{\rm MS}$ with an additional, momentum-dependent counter term which removes this unphysical large-momentum tail. Our renormalization condition is simple: the static two-point function should have the trivial $\boldsymbol{p}^2$-dependence in vacuum,
\begin{align}\label{eq:rc2}
    \Sigma_{\rm vac}^\pi(p_0 = 0, \boldsymbol{p}^2) \equiv Z \boldsymbol{p}^2 + \bar m^2\,.
\end{align}
The parameter $Z$ is fixed by the condition in \Eq{eq:Zcond} to $Z=1$.
This not only entails that the screening and curvature masses are identical, but, owing to Lorentz invariance in vacuum, they are also identical to the vacuum pole mass, i.e.\ $\bar m^{\rm vac} = m_s^{\rm vac} = m_p^{\rm vac}$. The renormalization conditions for the meson masses discussed above are hence more meaningful, as we automatically fix the physical pole masses and the propagators have unit residue at these pole. This \emph{vacuum-subtracted $\overline{\rm MS}$ scheme} is realized by considering the momentum dependence of the self-energy correction relative to the vacuum contribution,
%
\begin{align}\label{eq:two_point_re2}
\nonumber 
 \Pi^{\pi,\mathrm{vs}}_{\mathrm{vac}}(p^2;m_f)&=\Pi^{\pi,\mathrm{re}}_{\mathrm{vac}}(p^2;m_f)\\[2ex] \nonumber
 &-\Pi^{\pi,\mathrm{re}}_{\mathrm{vac}}(p^2_0=0,\boldsymbol{p}^2;m_f^{\mathrm{vac}})\\[2ex]
 &+\Pi^{\pi,\mathrm{re}}_{\mathrm{vac}}(0;m_f^{\mathrm{vac}})\,.
\end{align}
%
$m_f^{\mathrm{vac}}$ is the constituent quark mass at $T=0$ and $\mu=0$. This definition ensures that the unphysical large-momentum contribution is removed while the location of the moat regime (i.e.\, where $Z^\perp <0$) and the momentum-independent part of the two-point function remain unaffected. The result is shown in the red solid lines in \Fig{fig:renorm_compare}. From the right panel we see that while the moat regime remains, the two-point function is now a monotonically increasing, positive function at large $\boldsymbol{p}^2$. The vacuum subtraction introduces a momentum dependent component to the renormalization procedure, which can be interpreted as mimicking a renormalization scale running. This ensures the proper treatment of the two-point function over all momentum scales \footnote{We note that a similar result could presumably be achieved by using some form of Pauli-Villars regularization}. Naturally, the introduction of this additional momentum structure can alter the analytic structure of the two-point function.
We therefore use the artificial distinction between $p_0$ and $\boldsymbol{p}$ in the vacuum contribution here. This is to ensure that the vacuum subtraction does not change upon analytic continuation to Minkowski space. It would otherwise lead to artificial contributions to the spectral function. However, it still changes the analytic structure for complex spatial momenta.
This will be addressed in more detail in \cite{DM2}.

The discussion of this section shows that without a unique renormalization condition for the spatial wave function renormalization and a more accurate treatment of the momentum-dependence of the two-point function by going beyond RPA, statements about the moat regime can only be of qualitative nature. We hence focus on qualitative and structural aspects of the moat regime in the following.

%
\begin{figure}[t]
\includegraphics[width=0.49\textwidth]{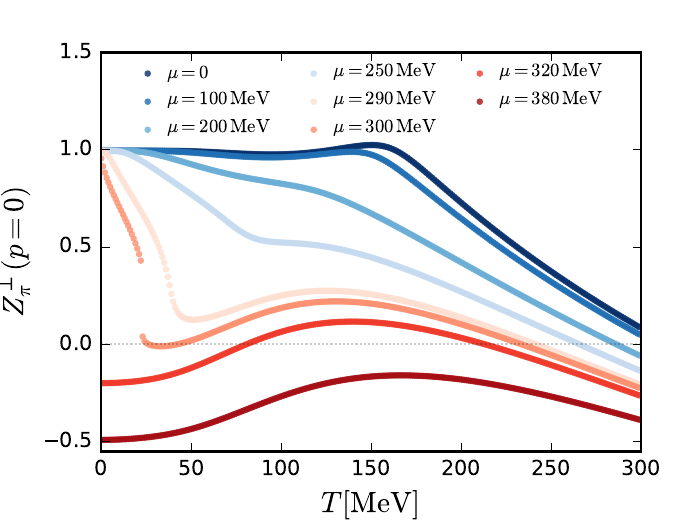}
\caption{Spatial pion wave function renormalization $Z_\pi^\perp(0)$ at vanishing momentum as function of temperature for various chemical potentials. }\label{fig:ZpiM_mu_T}
\end{figure}
%

\section{Shape of the moat regime}\label{sec:dissecting}

\subsection{Phase diagram and renormalization}\label{subsec:moat}

%
\begin{figure*}[t]
\includegraphics[width=0.49\textwidth]{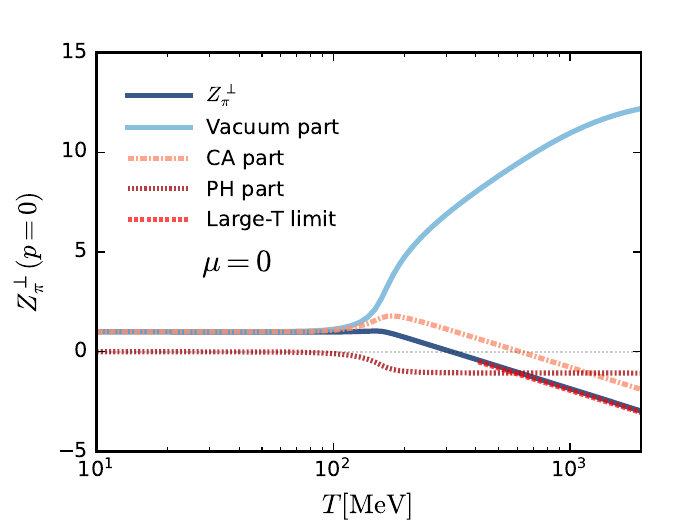}
\includegraphics[width=0.49\textwidth]{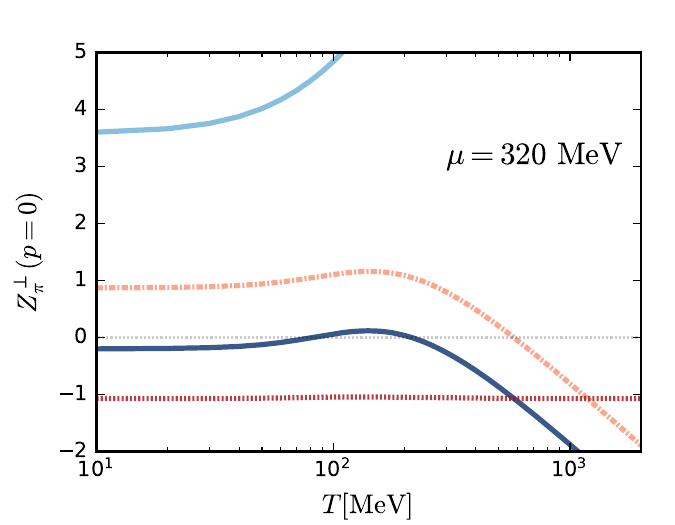}
\caption{Spatial pion wave function renormalization $Z_\pi^\perp(0)$ at large $T$, $\mu=0$ (left panel) and $\mu=320$ MeV (right panel). The dark blue solid lines are the full results. The dashed red line is the asymptotic behavior from the analytic large-$T$ limit in \Eq{eq:ZpilargeT}. The light blue solid, peach dot-dashed and brown dotted lines are the vacuum, creation-annihilation (CA) and particle-hole (PH) contributions respectively.}\label{fig:ZpiM_highT}
\end{figure*}
%

Here we study the moat regime in the phase diagram of the QM model. To this end, as in the previous section, we solve the gap equation \eq{eq:eom} to obtain the constituent quark mass at finite $T$ and $\mu$. $Z^\perp_\pi$ is then computed from \Eq{eq:Z_v_t} and the renormalization condition in \Eq{eq:Zcond}. 
In \Fig{fig:ZpiM_mu_T}, we show $Z^\perp_\pi$ as a function of temperature for different quark chemical potentials. Focusing on lower temperatures first, we see that $Z^\perp_\pi(T)$ is decreasing with increasing $\mu$. For $\mu \gtrsim \mu_{\rm CEP} \approx 290$\,MeV it becomes negative at small and intermediate $T$ and the system enters the moat regime. The jump seen at $\mu = 300$\, MeV reflects the first-order chiral phase transition. Importantly, the moat regime occurs adjacent to the phase boundary in the chirally restored phase.

We also find that at large $T$, and irrespective of $\mu$, $Z^\perp_\pi$ always turns negative. This has also been observed in Refs.\ \cite{Topfel:2024iop,Cao:2025zvh}, and can be understood analytically from the large-$T$ expansion of $Z^\perp_\pi$ in \Eq{eq:Zpi}. As we derive in \App{app:HighT}, at asymptotically large temperature and $\mu= 0$ the spatial pion wave function renormalization is
\begin{align}\label{eq:ZpilargeT}
    Z^\perp_\pi(0;T) \;\xrightarrow{T\rightarrow \infty}\; - \frac{h^2 N_c}{8\pi^2}\, \ln\bigg(\frac{T}{M}\bigg)\,.
\end{align}
Hence, $Z^\perp_\pi$ is negative and decreases logarithmically at large $T$ in RPA, even at $\mu=0$. This is shown in \Fig{fig:ZpiM_highT}. In general, it is an artifact of the present one loop approximation, and is mitigated once renormalization of the Yukawa coupling is taken into account as well. We will get back to this in \Sec{subsec:yukawa}.

\begin{figure*}[t]
\includegraphics[width=0.49\textwidth]{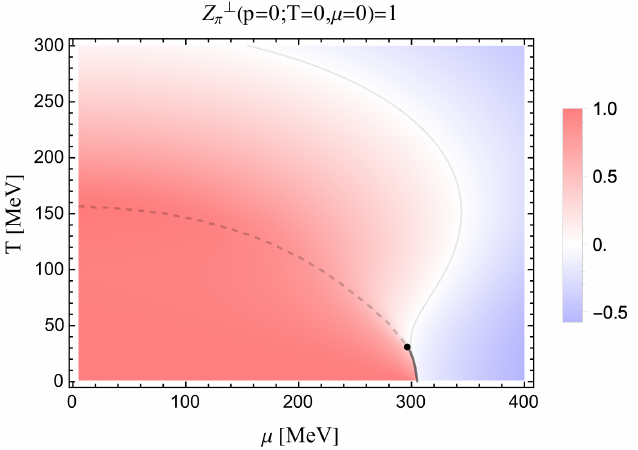}
\includegraphics[width=0.49\textwidth]{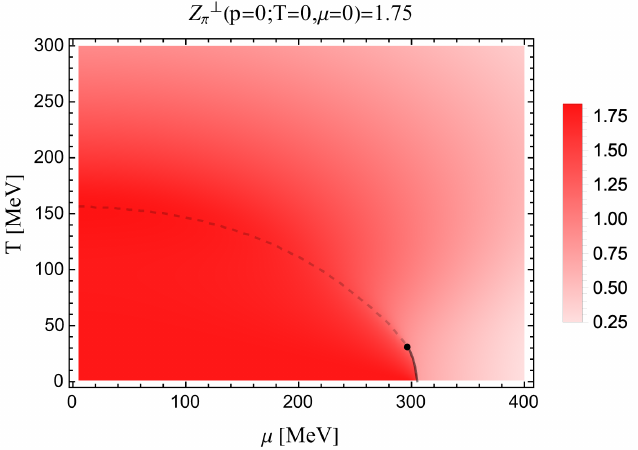}
\caption{Spatial pion wave function renormalization in the phase diagram. The left plot is the result of the renormalization condition in \Eq{eq:rc2} with $Z=1$ and on the right we used $Z=1.75$. As discussed in the text, the latter choice is unphysical here, so the right plot exemplifies the consequences of improper renormalization. The gray line indicates $Z^\perp_\pi=0$. The dashed line shows the pseudocritical temperature of the chiral crossover, which ends in a CEP, shown by the black dot, and then continues as a first-order transition along the solid black line.}\label{fig:phase}
\end{figure*}

To study the moat regime in the phase diagram more clearly, we show $Z^\perp_\pi$ in the entire phase diagram in \Fig{fig:phase}. To illustrate the importance of the renormalization condition,
we show the result using our renormalization condition in \Eq{eq:rc2} with $Z=1$ on the left and, for comparison, set $Z=1.75$ on the right. The latter corresponds to a change in renormalization scale without enforcing a proper renormalization condition, cf.\ \Fig{fig:ZpiM}. If the residue of the propagator is held fixed, any $Z \neq 1$ implies that curvature and pole masses are unequal in vacuum, and fixing the curvature masses based on experimental results, as we do here, becomes inaccurate. For example, with $Z=1.75$ and a pion curvature mass $\bar m_\pi^{\rm vac} = 136$\,MeV, the pole mass is $m_{p,\pi}^{\rm vac} = \bar m_\pi^{\rm vac}/\sqrt{Z} \approx 103\,{\rm MeV}$. Hence, the right plot of \Fig{fig:phase} is unphysical, and should only highlight the importance of proper renormalization \footnote{Note that the solution of the gap equation only depends on the curvature masses in RPA, so the chiral phase boundary is not affected by changes in $Z$}.

While for $Z=1$ there is a large moat regime in the chirally restored part of the phase diagram, there is no moat regime for any $T<300$\,MeV and $\mu<400$\,MeV for $Z=1.75$. The gray contour on the left plot indicates the boundary of the moat regime, $Z^\perp_\pi=0$. This contour starts at high temperature for low density, gradually shifts toward lower temperatures with increasing chemical potential and finally merges with the first order phase boundary. This shape of the moat regime is consistent with that found in the recent calculations of QM models \cite{Topfel:2024iop, Cao:2025zvh}. 

Interestingly, the boundary of the moat regime roughly follows the chiral transition, with an offset that vanishes around the CEP. The underlying reason is that the contributions to the quartic meson coupling $\lambda$ and the spatial wave function renormalization $Z^\perp$ from the functional quark determinant are identical \cite{Nickel:2009ke}. This leads to a coincidence of the CEP with a possible Lifshitz point if mesonic corrections are ignored \cite{Carignano:2014jla}, and more generally to a direct link between the moat regime and the first-order chiral transition, where the quartic meson coupling is negative. In the present case, neither a Lifshitz point nor an inhomogeneous phase occurs, but a link between the chiral transition and the moat regime still remains from the quark determinant. The overall shape of $Z^\perp_\pi$ in the phase diagram remains unchanged in the right plot, but its overall value has increased by a constant value of $0.75$, moving the moat regime to much higher temperatures and densities. Note that the chiral phase boundary is the same in both cases, as the wave function renormalization does not feed back into the effective potential in RPA.

%
%

Following our results in Ref.\ \cite{Fu:2024rto}, we can split the thermal contributions to the wave function renormalization into those from relativistic particle–antiparticle creation and annihilation (CA) processes, and those from non-relativistic particle–hole (PH) fluctuations related to Landau damping. The corresponding equations are given in \App{app:thre_fun}. 
We show these contributions at $\mu = 0$ in the left panel of \Fig{fig:ZpiM_highT}.
Below $T\approx 100$\,MeV, the values of both the CA and PH parts remain zero, and at higher temperatures negative values are excited. In the high temperature limit, the PH part approaches a constant, while the CA part keeps decreasing as the temperature rises. The vacuum contribution, which is part of the CA contribution, is always positive. We hence find that the large-$T$ behavior in RPA, see \Eq{eq:ZpilargeT}, is triggered by CA processes. In contrast, the moat regime at small and intermediate $T$ seen in \Fig{fig:ZpiM_mu_T} and \Fig{fig:phase} is solely due to PH processes. This is demonstrated explicitly in the right panel of \Fig{fig:ZpiM_highT}, where we show $Z_\pi^\perp$ at $\mu = 320$\, MeV. 
$Z^\perp_\pi$ is driven to negative values by PH fluctuations at low temperatures, returns to positive values at $T\simeq20\,\mathrm{MeV}$ and is subsequently driven negative again by CA processes.
In the left plot of \Fig{fig:phase}, the moat regime in the bottom right corner is due to PH fluctuations, while the turning point of the $Z_\pi^\perp = 0$ contour at $T \gtrsim 150$\,MeV can be attributed to CA processes. 
We hence corroborate the results of Ref.\ \cite{Fu:2024rto}, where it was found that the moat regime in QCD arises from PH processes. Since the negative $Z_\pi^\perp$ at large $T$ has a different physical origin, one shall perhaps not classify it as a moat regime.  

Note that the shape of the moat regime found here and in other low-energy models is completely different from that obtained in first-principles QCD calculations in \cite{Fu:2019hdw,Fu:2024rto}. In QCD, the dominance of the negative CA contribution at large $T$ has not been observed, and the moat regime found in \cite{Fu:2019hdw,Fu:2024rto} is only due to PH processes. As we will demonstrate next, the differences between QCD and our model calculation lies to a large extent in the treatment of the interaction between quarks and mesons.

%
\begin{figure}[b]
\includegraphics[width=0.49\textwidth]{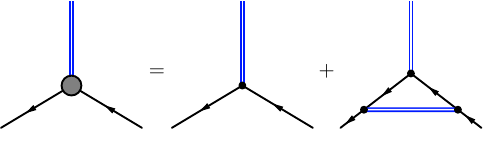}
\caption{The quark-meson interaction at one-loop. The gray circle denotes the renormalized quark-meson vertex, while the black dot is the bare coupling. We use free propagators with $T$ and $\mu$ dependent masses determined in mean-field approximation for the internal lines.}\label{fig:yukawa_loop}
\end{figure}
%

Before we move on, it is important to emphasize that we do not find an inhomogeneous instability within the moat regime: while the pion two-point function can have a minimum at nonzero spatial momentum, it is always larger than zero. This overlaps with the conclusion of Ref.\ \cite{Pannullo:2024sov} that, while the existence of an inhomogeneous instability in the $(3\!+\!1)$-dimensional Gross-Neveu model is highly regularization scheme and scale dependent, the moat regime is a much more robust feature. The absence of an inhomogeneous instability here can presumably be attributed to the relatively light sigma meson mass we have chosen as a renormalization condition \cite{Carignano:2014jla}. We add that any renormalization scale dependence is completely removed once the self-energy corrections are properly renormalized.

\subsection{In-medium Yukawa coupling}\label{subsec:yukawa}

%
\begin{figure*}[t]
\includegraphics[width=0.47\textwidth]{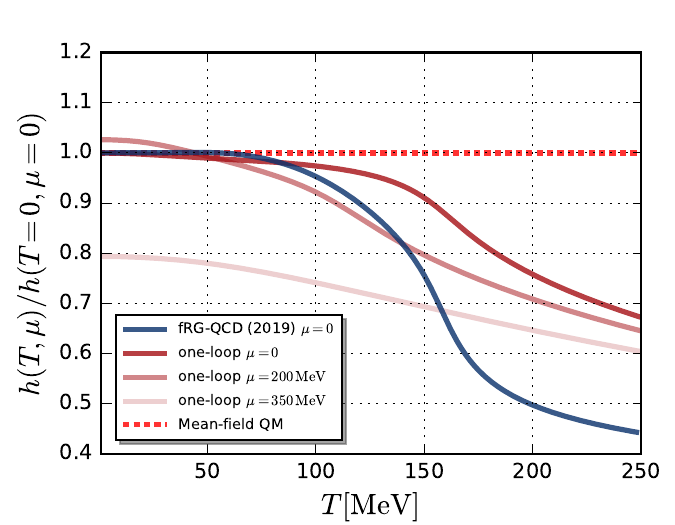}
\includegraphics[width=0.51\textwidth]{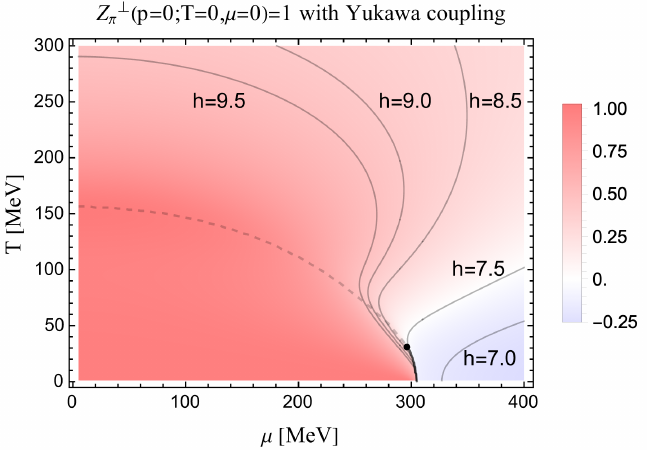}
\caption{\emph{Left:} Comparison of the pion Yukawa coupling as functions of temperature from FRG-QCD (blue solid line) \cite{Fu:2019hdw}, from our one-loop calculation in \Fig{fig:yukawa_loop} (dark red solid line) and in the mean-field approximation at vanishing density (red dashed line). We also show examples for the one-loop Yukawa coupling at nonzero $\mu$. \emph{Right:} The spatial pion wave function renormalization in the phase diagram computed with the one-loop Yukawa coupling from \Fig{fig:yukawa_loop}. The heat map is for a value of $h=7.5$ in vacuum. The solid gray lines show the boundaries of the moat regime, $Z^\perp_\pi=0$, for different vacuum values of $h$.}\label{fig:hT}
\end{figure*}
%

As seen from the self-energy in \Eq{eq:pion-two-point} and the wave function renormalization in \Eq{eq:Zpi}, the interaction between quarks and mesons play a crucial role for the correlation of mesons. The strength of the quark-meson interaction is determined by the Yukawa coupling $h$. In RPA, the Yukawa coupling is treated as constant and does not change with neither temperature nor density. This clearly is not true in general, and can potentially have large effects on the phase diagram. In particular given that $h$ directly controls the strength of CA and PH contributions.

As we have shown in the previous section, owing to the CA contributions, there always is a moat regime at large $T$ for any $\mu$ in the QM model in RPA. This is in contrast to QCD, where the moat regime only occurs at $\mu \gtrsim 150$\,MeV \cite{Fu:2019hdw,Fu:2024rto}. As we will demonstrate now, this considerable qualitative difference is largely due to in-medium modifications of the Yukawa coupling, which are absent in RPA. To this end, we compute the one-loop correction to the Yukawa coupling as shown in \Fig{fig:yukawa_loop}. The details of the computation are given in \App{app:yuk}.

In the left panel of \Fig{fig:hT} we compare the resulting Yukawa coupling to its mean-field value and the QCD result obtained with the FRG \cite{Fu:2019hdw} as functions of temperature at vanishing chemical potential. While the mean-field coupling is constant, both the one-loop and the QCD coupling decrease with increasing temperature due to in-medium screening of the quark-antiquark potential.
The small interaction strength at large $T$ will suppress the self-energy correction of the meson two-point correlation function in \Eq{eq:pion-two-point}.

To demonstrate this effect, we replace the constant coupling $h$ in the spatial pion wave function renormalization in \Eq{eq:Zpi} by the temperature and chemical potential dependent coupling obtained from \Fig{fig:yukawa_loop} and \App{app:yuk}. In general, the Yukawa coupling enters both through the vertices in \Fig{fig:feynman-dia-meson} and the quark mass $m_f^2 = h^2 \rho_0/2$. The latter is already $T$ and $\mu$ dependent in mean-field, because $\rho_0$ is the solution of the equation of motion \eq{eq:eom}. The in-medium modification of $h$ only give small corrections there. In contrast, these corrections have a large effect when taken into account in the vertices in \Fig{fig:feynman-dia-meson}. This is seen in the right panel of \Fig{fig:hT}, where we show $Z_\pi^\perp$ with the one-loop Yukawa vertex in the phase diagram. We see that the smaller the Yukawa coupling is in vacuum, the more the moat regime is moved towards larger $\mu$ in the phase diagram. For $h(T=0,\mu=0) = 7.5$, which corresponds to a realistic constituent quark mass of $m_f \approx 350$\,MeV \cite{Cyrol:2017ewj}, the location of the moat regime relative to the location of the CEP is qualitatively similar to QCD \cite{Fu:2019hdw,Fu:2024rto}. The larger the vacuum coupling, the smaller the suppression of CA processes, so for $h(T=0,\mu=0) \gtrsim 9$ the moat regime is similar to the pure RPA result in \Fig{fig:phase}. Conversely, for $h(T=0,\mu=0) \lesssim 6.5$, the moat regime disappears completely since also PH fluctuations are suppressed with decreasing Yukawa coupling.

The present analysis, using an \emph{ad hoc} procedure of feeding the $T$ and $\mu$ dependence of the one-loop coupling into the one-loop self-energy, should only be taken as indicative, not conclusive. Without a self-consistent (higher-loop) calculation of the vertex corrections, and taking them into account also in the gap equation, chiral Ward identities may be violated \footnote{We thank Michael Buballa for pointing this out.}. This is evident, e.g., from the Golberger-Treiman relation in the present case, $h f_\pi = m_f$. Still, our results show that in-medium modifications of quark-meson interactions have a significant effect on the moat regime. We expect that, in qualitative agreement with our results, an actual higher-loop calculation of the pion self-energy in the low-energy model will lead to a moat regime that is absent at low $\mu$, and located in the vicinity of the CEP and the first-order chiral phase transition. This is supported by the FRG results in \cite{Fu:2019hdw, Fu:2024rto}, where all this is done self-consistently in QCD.

\section{Summary and conclusion}\label{sec:summary}

The moat regime has recently emerged as a characteristic feature of spatial modulations in various systems, including QCD. In order to understand how to systematically investigate this regime using effective models, we considered a two-flavor QM model that captures key features related to the chiral phase transition of QCD. Unlike in QCD, the relevant low-energy degrees of freedom are not emergent in such a model, but put in by hand. It is crucial to understand how to properly renormalize the system in this case, as otherwise reliable statements about its properties are obstructed by uncontrolled renormalization scale and scheme dependencies.

In low-energy models of QCD, one typically adjusts the model parameters in order to reproduce meson masses and decay constants, see, e.g., \cite{Klevansky:1992qe}. This implicitly fixes counter terms for the renormalization of the effective potential. For the moat regime, including inhomogeneous phases, however, the momentum dependence of correlation function plays a crucial role. We have pointed out in this work that the procedure described above is insufficient, and additional renormalization conditions are required to extract meaningful and robust information on the moat regime.

This is apparent in the QM model, or any Yukawa and NJL-type theory in three or more spatial dimensions, as the momentum-dependent part of the boson self-energy is not renormalized by the counter terms for the effective potential and hence requires additional regularization. This gives rise to a nontrivial wave function renormalization $Z$ of the theory. One can therefore consider $Z$ as an additional parameter of the model, which needs to be fixed through an appropriate renormalization condition. Note that this is expected from standard renormalization theory, as the wavefunction renormalization is power-counting marginal. Since the spatial contribution $Z^\perp$ determines the moat regime, it is evident that without fixing this parameter, results on the moat regime can become highly scheme and scale-dependent, see, e.g., \cite{Partyka:2008sv, Buballa:2020nsi, Pannullo:2023cat, Pannullo:2024sov}. While we have shown that the renormalization scale dependence of the moat regime is completely removed if $Z$ is renormalized, we did not thoroughly investigate the possibility of a remaining regularization scheme dependence. And even though we put forward a convenient renormalization scheme where pole and screening masses are identical in vacuum, a more direct condition that fixes $Z^\perp$, for example, as we suggested, by using screening masses measured on the lattice at finite temperature, could have its advantages in model studies. In particular since certain choices for $Z^\perp$ might obscure the identification of chiral partners in the chirally restored phase. We will get back to this in a forthcoming work \cite{DM2}. For a discussion of RG consistency in the context of the FRG we refer to Ref.\ \cite{Braun:2018svj}. 

Furthermore, even with a renormalization condition for $Z^\perp$, minimal subtraction of divergent contributions plus some constants that enforce the renormalization conditions, the meson two-point function becomes unphysical at large spatial momenta in RPA: It eventually turns negative, suggesting an instability towards an inhomogeneous phase already at $\mu=0$. We therefore proposed the vacuum-subtracted $\overline{\rm MS}$ scheme, which enforces a trivial meson propagator in vacuum and removes the negative contribution at large $\boldsymbol{p}^2$, while leaving the location of the moat regime unaffected. Conventional on-shell renormalization can hence easily be implemented. However, this necessitates a momentum-dependent counter term which can potentially affect the analytic structure of correlation functions. We will discuss this in more detail in \cite{DM2}.

Applying these technical developments to the phase diagram, we find a large moat regime in the $(T,\mu_B)$ plane. In line with the results in Refs.\ \cite{Topfel:2024iop,Cao:2025zvh}, we do not find indications for an inhomogeneous instability, so a second-order transition to an inhomogeneous phase seems unlikely. We emphasize, however, that this crucially depends to the scalar meson mass $\bar m_\sigma$ \cite{Carignano:2014jla}. We have chosen a curvature mass $\bar m_\sigma^{\rm vac} = 480$\,MeV, in line with the QCD results in Ref.\ \cite{Fu:2019hdw}. This is consistent with the results of \cite{Carignano:2014jla}, where an inhomogeneous phase in the QM model is only found for relatively heavy scalar mesons, $\bar m_\sigma^{\rm vac} \gtrsim 590$\,MeV. Studying this mass dependence in light of the developments in this work would be a worthwhile task.

We also confirmed the findings of Refs.\ \cite{Topfel:2024iop,Cao:2025zvh}, where a moat regime seems to occur for any $\mu\geq 0$ at large $T$ in the RPA QM model. We have shown analytically that creation-annihilation processes arising in the fermion determinant will always lead to a negative contribution to $Z^\perp$ and are hence responsible for this phenomenon. In contrast, the actual moat regime is triggered by particle-hole fluctuations and only occurs at sufficiently large $\mu$, see also Ref.\ \cite{Fu:2024rto}. We have demonstrated that the CA-induced moat behavior is an artifact of the constant quark-meson coupling in RPA. If in-medium modifications of this interaction are taken into account, the moat regime is only found in the vicinity of the CEP and the first order chiral transition. Note that the connection between the moat regime and the CEP/first-order line follows from the fact that the contribution from the quark determinant to the quartic meson coupling, whose sign change indicates a first-order transition, and $Z^\perp$ are identical \cite{Nickel:2009ke}.

We believe that our results form the basis for systematic studies of the moat regime in effective models. Their utility goes beyond the QCD context, as the moat regime can help to shed light on various systems involving spatial modulations \cite{Pisarski:2021qof, Rennecke:2021ovl, Nussinov:2024erh}. First applications will be presented in Ref.\ \cite{DM2}.

\vspace{1ex}
\section{Acknowledgements}
We thank the members of fQCD collaboration \cite{fQCD}, especially Jens Braun, Wei-jie Fu and Jan Pawlowski, as well as Michael Buballa, Zohar Nussinov, Mike Ogilvie, Laurin Pannullo, Rob Pisarski, Stella Schindler, Lorenz von Smekal and Marc Winstel for discussions and comments. This work is supported by the Deutsche Forschungsgemeinschaft (DFG, German Research Foundation) through the CRC-TR 211 ``Strong-interaction matter under extreme conditions'' -- project number 315477589 -- TRR 211. S.\ Y.\ is supported by the Alexander von Humboldt foundation.
\section{Data availability}
The data are not publicly available. The data are available from the authors upon reasonable request.

\appendix

\titleformat{\section}[block]{\small\bfseries\filcenter}{Appendix \Alph{section}:}{1em}{}
\renewcommand{\thesection}{\Alph{section}}
\renewcommand{\theequation}{\Alph{section}\arabic{equation}}

\section{Threshold functions}\label{app:thre_fun}

Here we provide the explicit expressions for the fermion loop functions, which are used in the equations for the pion two-point function \labelcref{eq:pion-two-point} and the wave function renormalization \labelcref{eq:Zpi}. The threshold function of $n$-th order at vanishing external momentum is given by
%
\begin{align}
\mathcal{F}_{(n)}(q)=T\sum_n \bar{G}^n_{f}(q,m^2_f;T,\mu)\,,
\end{align}
%
where we use the scalar part of the quark propagator $\bar{G}_f(q,m^2_f;T,\mu)=1/((q_0+i\mu)^2+\boldsymbol{q}^2+m^2_f)$. We only need the lowest three orders here (and often omit the momentum arguments in the following for the sake of brevity),
%
\begin{align}\label{eq:threshold_fun}
\mathcal{F}_{(1)}&=\mathcal{F}^{\mathrm{vac}}_{(1)}+\frac{1}{2E_q}\bigg[-n_F(E_q;T,\mu)-n_F(E_q;T,-\mu)\bigg]\,,\\[2ex]
\mathcal{F}_{(2)}&=\mathcal{F}^{\mathrm{CA}}_{(2)}+\mathcal{F}^{\mathrm{PH}}_{(2)}\,,\\[2ex]
\mathcal{F}_{(3)}&=\mathcal{F}^{\mathrm{CA}}_{(3)}+\mathcal{F}^{\mathrm{PH}}_{(3)}\,.
\end{align}
%
The vacuum part is given by
%
\begin{align}
\mathcal{F}^{\mathrm{vac}}_{(1)}=\frac{1}{2E_q}\,.
\end{align}
%
We split the threshold functions of second and third order into two parts,
%
\begin{align}\label{eq:threshold_fun_f2}
\mathcal{F}^{\mathrm{CA}}_{(2)}&=\mathcal{F}^{\mathrm{vac}}_{(2)}+\frac{1}{4E_q^{3}}\bigg[-n_F(E_q;T,\mu)-n_F(E_q;T,-\mu)\bigg]\,,\nonumber\\[2ex]
\mathcal{F}^{\mathrm{PH}}_{(2)}&=\frac{1}{4E_q^{3}}\bigg[E_q\Big(n_F'(E_q;T,\mu)+n_F'(E_q;T,-\mu)\Big)\bigg]\,,
\end{align}
%
%
\begin{align}\label{eq:threshold_fun_f3}
\mathcal{F}^{\mathrm{CA}}_{(3)}=&\mathcal{F}^{\mathrm{vac}}_{(3)}+\frac{3}{16E_q^{5}}\bigg[-n_F(E_q;T,\mu)-n_F(E_q;T,-\mu)\bigg]\,,\nonumber\\[2ex]
\mathcal{F}^{\mathrm{PH}}_{(3)}=&\frac{1}{16E_q^{5}}\Bigg\{E_q^2\Big(-n_F''(E_q;T,\mu)-n_F''(E_q;T,-\mu)\Big)\nonumber\\[2ex]
&+3\bigg[E_q\Big(n_F'(E_q;T,\mu)+n_F'(E_q;T,-\mu)\Big)\bigg]\Bigg\}\,.
\end{align}
%
`CA' stands for the particle creation and annihilation in the quark loop and `PH' for particle-hole fluctuations. The derivatives of the fermion distribution are taken with respect to the energy,
%
\begin{align}
n_F^{(n)}(x;T,\mu)=\frac{\partial^n n_F(x;T,\mu)}{\partial x^n}\,.
\end{align}
%
Their vacuum parts can be given by
%
\begin{align}
\mathcal{F}^{\mathrm{vac}}_{(2)}&=\frac{1}{4E_q^3}\,,\\[2ex]
\mathcal{F}^{\mathrm{vac}}_{(3)}&=\frac{3}{16E_q^5}\,.
\end{align}
%
In addition, for the momentum dependent pion two-point function, we need the threshold function at finite external momentum,
%
\begin{align}
\begin{split}
&\mathcal{FF}^-_{(n,m)}(p,q)\\[2ex]
&\quad= T\sum_n \bar{G}^n_{f}(q,m^2_f;T,\mu)\bar{G}^m_{f}(q-p,m^2_f;T,\mu)\,.
\end{split}
\end{align}
%
We only need the lowest order, $n=1$ and $m=1$,
%
\begin{align}\label{eq:threshold_fun_p}
\mathcal{FF}^-_{(1,1)}&(p_0,\boldsymbol{p},\boldsymbol{q};m_f,T,\mu)=\frac{1}{4E_qE_{q-p}}\nonumber\\[2ex]
\times\Bigg\{&\frac{n_F\big(E_{q-p};T,\mu\big)+n_F\big(E_q;T,-\mu\big)}{i p_0-E_q-E_{q-p}}\nonumber\\[2ex]
+&\frac{-n_F(E_q;T,\mu)-n_F(E_{q-p};T,-\mu)}{i p_0+E_q+E_{q-p}}\nonumber\\[2ex]
+&\frac{-n_F\big(E_q;T,-\mu \big)+n_F\big(E_{q-p};T,-\mu\big)}{i p_0-E_q+E_{q-p}}\nonumber\\[2ex]
+&\frac{n_F\big(E_q;T,\mu\big)-n_F\big(E_{q-p};T,\mu\big)}{i p_0+E_q-E_{q-p}}\Bigg\}\nonumber\\[2ex]
+&\mathcal{FF}^{-,\mathrm{vac}}_{(1,1)}\,.
\end{align}
%
The minus in the superscript denotes the sign of the external spatial momentum, but we emphasize that the results are independent of the momentum rooting. The quark energy is $E_q=\sqrt{\boldsymbol{q}^2+m^2_f}$. The vacuum part of the function is
%
\begin{align}
\mathcal{FF}^{-,\mathrm{vac}}_{(1,1)}&=\frac{1}{4E_qE_{q-p}}\nonumber\\[2ex]
&\times\Bigg\{\frac{1}{ip_0-E_q-E_{q-p}}+\frac{1}{ip_0+E_q+E_{q-p}}\Bigg\}\,.
\end{align}
%
Note that the regularization and renormalization of all the vacuum parts of the threshold functions are introduced in \Sec{sec:regularization}.

\section{Yukawa coupling}\label{app:yuk}

Here we provide details on the quark-pion Yukawa coupling shown in \Fig{fig:yukawa_loop}. Since we are focusing on in-medium modifications, we do not take the full momentum dependence of the coupling into account. Feeding this into the vertex in \Fig{fig:yukawa_loop} would be equivalent to evaluating a higher-loop diagram. For simplicity, we will not do this here, but rather use the one-loop coupling at a fixed momentum configuration. Given our findings in \Sec{subsec:yukawa}, this turns out to be sufficient for the present purposes. We hence choose to evaluate this diagram at zero external spatial momentum for the pion and the quarks. The external pion frequency is also set to zero (i.e.\ the lowest bosonic Matsubara mode) and the external quark frequency is set to the lowest fermionc Matsubara mode, $p_0=\pi T$. The one-loop diagram in \Fig{fig:yukawa_loop} then yields for the temperature and chemical potential dependent Yukawa coupling:
%
\begin{align}
h_{\pi}(T,\mu)=-\frac{h^3}{4N_f\,\pi^2}\bigg[&(N_f^2-1)\,I_{(1,1)}(m^2_\pi,m^2_f;T,\mu;\pi T)\nonumber\\[2ex]
&-I_{(1,1)}(m^2_\sigma,m^2_f;T,\mu;\pi T)\bigg]
\end{align}
%
The boson–fermion mixed loop function can be expressed as 
%
\begin{align}
I_{(1,1)}(m^2_\phi,&m^2_f;T,\mu;p_0)\nonumber\\[2ex]
&=\int dq\,q^{d-1}\,\mathcal{FB}_{(1,1)}(m^2_\phi,m^2_f;T,\mu;p_0)\,,
\end{align}
%
with 
%
\begin{align}
\mathcal{FB}_{(n,m)}(q)=T\sum_{n_q} \bar{G}^n_{f}(q,m^2_f;T,\mu)\bar{G}^m_{b}(q,m^2_b;T)\,,
\end{align}
%
where the boson propagator is $\bar G_b(q,m^2_b;T)=1/(q_0^2+\boldsymbol{q}^2+m_b^2)$. Again, we only need the lowest order here,
%
\begin{align}\label{eq:FB11}
&\mathcal{FB}_{(1,1)}=\nonumber\\
&\quad\frac{1}{2}\,\mathrm{Re}\bigg\{-n_B(E_b;T)\frac{1}{E_b}\frac{1}{(ip_0-\mu+E_b)^2-E_q^2}\nonumber\\[2ex]
&\quad-\big(n_B(E_b;T)+1\big)\frac{1}{E_b}\frac{1}{\big(ip_0-\mu-E_b\big)^2-E_q^2}\nonumber\\[2ex]
&\quad+n_F(E_q;T,-\mu)\frac{1}{E_q}\frac{1}{\big(ip_0-\mu-E_q\big)^2-E_b^2}\nonumber\\[2ex]
&\quad+\big(n_F(E_q;T,\mu)-1\big)\frac{1}{E_q}\frac{1}{\big(ip_0-\mu+E_q\big)^2-E_b^2}\bigg\}\,,
\end{align}
%
with the Bose-Einstein distribution $n_B(x;T)=1/\big(\mathrm{exp}(x/T)-1\big)$. The free boson energy is $E_b=\sqrt{\boldsymbol{q}^2+m_b^2}$. 

We close this section with a comment on the choice of the external fermion frequency $p_0$.
Since we do not compute full higher-loop corrections of the self-energy that encode the vertex corrections, but rather just input the one-loop coupling at a fixed momentum configuration, some caution is advised. The reason is that all correlation function should respect the Silver-Blaze property \cite{Cohen:2003kd}. This entails that at $T=0$ and chemical potentials below the density onset ($\mu \lesssim 300$\,MeV in our case), the chemical potential dependence of any correlation function is completely described by a shift $p_{0j} + i \alpha_j \mu$ of the frequencies of the external legs, where $\alpha_j$ is the quark number of the $j$-th leg \cite{Marko:2014hea, Khan:2015puu,Fu:2016tey}.

As is evident from \Eq{eq:FB11}, where $p_0$ is the external quark momentum, the one-loop Yukawa coupling is consistent with the Silver-Blaze property. However, we input this coupling into the pion self-energy, which, according to Silver-Blaze, should not depend on $\mu$ at $T=0$ below the density onset. If we want to make sure that our choice of $p_0$ for $h_\pi$ does not spoil this, a natural choice would be $p_0=\pi\,T-i\mu$. We have checked that this does not make a qualitative difference to $p_0=\pi\,T$ here, because, first, the $\mu$-dependence of $h_\pi$ below onset is very small anyway, see \Fig{fig:hT}, and second, the moat regime occurs above the density onset. In addition, we also have to take the real part, since the Yukawa coupling for fixed $p_0$ is complex, see the discussion in Ref.\ \cite{Pawlowski:2014zaa}. We emphasize that all this would not be necessary if we directly computed the vertex corrections to the self-energy in terms of a higher-loop diagram, see the discussion in Ref.\ \cite{Fu:2016tey}. But since we are mainly interested in the $T$ and $\mu$-dependent corrections to $h_\pi$ we stick to our more pragmatic choice here.

\section{High-temperature expansion}\label{app:HighT}

Here we derive the high-temperature expansion of the spatial pion wave function renormalization discussed in \Sec{subsec:moat}. \Eq{eq:Zpi} can be written as
%
\begin{align}
\begin{split}
&Z^\perp_\pi(T)\\[2ex]
&\quad =1-\frac{h^2N_c}{\pi^2}\bigg[-I_2(m,T)+\frac{2}{3}\tilde I_3(m,T)\bigg]
\,.
\label{eq:ZpiA}
\end{split}
\end{align}
%
We focus on $\mu=0$ and consider each of the last two terms separately, starting from the first one,
\begin{align}
 I_2(m,T) &= \int\! dq\,q^{d-1} \mathcal{F}_{(2)}(q^2) =  -\partial_{m^2} I_1(m,T)\,,\nonumber\\[2ex]
    I_1(m,T) &= \int\!dq\,q^{d-1}\,\mathcal{F}_{(1)}(q^2)\nonumber\\[2ex]
    &= \sum_{n=-\infty}^\infty \int\!dq\,q^{d-1}\, \frac{T}{\nu_n^2+q^2+m^2}\,,
    \label{eq:I2}
\end{align}
with $\nu_n = (2 n +1)\pi T$.
We work in general spatial dimension $d$ in order to apply dimensional regularization below. By expanding around $m^2 = 0$, we can carry out the momentum integral,
\begin{align}
    &I_1(m,T)\nonumber\\[2ex]
    &\quad= \sum_{n=-\infty}^\infty \sum_{l=0}^\infty (-1)^l m^{2l} \int\!dq\, \, \frac{q^{d-1} T}{(\nu_n^2+q^2)^{l+1}}\nonumber\\[2ex]
    &\quad=\sum_{l=0}^\infty (-1)^l m^{2l}\, \frac{\Gamma\big(l\!+\!1-\!\frac{d}{2}\big)\Gamma\big(\frac{d}{2}\big)}{2 \Gamma(l+1)}\! \sum_{n=-\infty}^\infty  \frac{T}{\nu_n^{2l+2-d}}\,,
\end{align}
where $\Gamma(z)$ is the $\Gamma$-function. The Matsubara sum over $n$ can be carried out using the $\zeta$-function, leading to
\begin{align}
  I_1(m,T) = \sum_{l=0}^\infty& (-1)^l m^{2l}\, \frac{\Gamma\big(l\!+\!1-\!\frac{d}{2}\big)\Gamma\big(\frac{d}{2}\big)}{2 \Gamma(l+1)}\nonumber\\[2ex]
  &\times\frac{2 T \big(2^{2l+2-d}-1\big)}{(2\pi T)^{2l+2-d}}\zeta(2l+2-d)
\end{align}
We now set $d=3-2\epsilon$ and expand around $\epsilon = 0$ for $l= 0,1,2$, noting that larger $l$ lead to terms that are suppressed at $T\rightarrow\infty$. This yields
\begin{align}
   I_1(m,T) &= -\frac{\pi^2 T^2}{12}\nonumber\\[2ex]
   &\quad-\frac{m^2}{4}\bigg[\frac{1}{2\epsilon}-\ln\bigg(\frac{T}{M}\bigg) + \gamma_E -1 -\ln\bigg(\!\frac{\pi}{2}\!\bigg)  \bigg]\nonumber\\[2ex]
   &\quad + \frac{7 m^4 \zeta(3)}{64\pi^2 T^2} + \mathcal{O}\bigg(\frac{1}{T^4}\bigg)
   + \mathcal{O}(\epsilon)
\end{align}
Inserting this into \Eq{eq:I2}, we get
\begin{align}
   I_2(m,T) &=  \frac{1}{4}\bigg[\frac{1}{2\epsilon}-\ln\bigg(\frac{T}{M}\bigg) + \gamma_E -1 -\ln\bigg(\!\frac{\pi}{2}\!\bigg)  \bigg]\nonumber\\[2ex]
   &\quad+ \mathcal{O}\bigg(\frac{1}{T^2}\bigg)+ \mathcal{O}(\epsilon)\,.
   \label{eq:I2fin}
\end{align}
For the last term of \Eq{eq:ZpiA} we need to evaluate
\begin{align}
 \tilde I_3(m,T) &= \int\! dq\,q^{d-1} q^2\mathcal{F}_{(3)}(q^2) =  \frac{1}{2}\partial_{m^2}^2 \tilde I_1(m,T)\,,\nonumber\\[2ex]
    \tilde I_1(m,T) &= \int\!dq\,q^{d+1}\,\mathcal{F}_{(1)}(q^2)\nonumber\\[2ex]
    &= \sum_{n=-\infty}^\infty \int\!dq\,q^{d+1}\, \frac{T}{\nu_n^2+q^2+m^2}\,.
    \label{eq:I3}
\end{align}
A calculation in full analogy to the one of $I_1(m,T)$ gives
\begin{align}
   \tilde I_1(m,T) &=
   \sum_{l=0}^\infty (-1)^l m^{2l}\, \frac{\Gamma\big(l\!-\!\frac{d}{2}\big)\Gamma\big(1+\frac{d}{2}\big)}{2 \Gamma(l+1)}\nonumber\\[2ex]
  &\quad\phantom{\sum_{l=0}^\infty}\times\frac{2 T \big(2^{2l-d}-1\big)}{(2\pi T)^{2l-d}}\zeta(2l-d)\nonumber\\[2ex]
   &=-\frac{7\pi^4 T^4}{120} + \frac{m^2 \pi^2 T^2}{8}\nonumber\\[2ex]
   &\quad+\frac{3 m^4}{16}\bigg[\frac{1}{2\epsilon}-\ln\bigg(\frac{T}{M}\bigg) + \gamma_E -\frac{4}{3} -\ln\bigg(\!\frac{\pi}{2}\!\bigg)  \bigg]\nonumber\\[2ex]
   &\quad+ \mathcal{O}\bigg(\frac{1}{T^2}\bigg)+ \mathcal{O}(\epsilon)\,.
\end{align}
Plugging this into \Eq{eq:I3} gives
\begin{align}
    \tilde I_3 &= \frac{3}{16}\bigg[\frac{1}{2\epsilon}-\ln\bigg(\frac{T}{M}\bigg) + \gamma_E -\frac{4}{3} -\ln\bigg(\!\frac{\pi}{2}\!\bigg)  \bigg]\nonumber\\[2ex]
    &\quad+ \mathcal{O}\bigg(\frac{1}{T^2}\bigg)+ \mathcal{O}(\epsilon)\,,
\end{align}
and with Eqs.\ \eq{eq:I2fin} and \eq{eq:ZpiA} we finally arrive at the large-$T$ behavior of the spatial wave function renormalization,
\begin{align}
    &Z_\pi^\perp(T)\nonumber\\[2ex]
    &\quad = 1 + \frac{h^2 N_c}{8\pi^2} \bigg[\frac{1}{2 \epsilon} - \ln\bigg(\frac{T}{M}\bigg) + \gamma_E - \frac{4}{3} - \ln\bigg(\frac{\pi}{2}\bigg) \bigg]\nonumber\\[2ex]
    &\qquad+ \mathcal{O}\bigg(\frac{1}{T^2}\bigg) + \mathcal{O}(\epsilon)\,.
    \label{eq:ZpilT}
\end{align}
The divergent term is, as expected, canceled exactly by the counter term in \Eq{eq:delZ}. We can hence set $\epsilon = 0$ after renormalization and conclude that $Z_\pi^\perp$ indeed becomes increasingly negative at asymptotically large $T$,
\begin{align}
  Z_\pi^\perp(T) \;\xrightarrow{T\rightarrow \infty}\; - \frac{h^2 N_c}{ 8\pi^2}\, \ln\bigg(\frac{T}{M}\bigg)\,.
\end{align}
\vfill

\bibliography{ref-lib}

\begin{thebibliography}{83}%
\makeatletter
\providecommand \@ifxundefined [1]{%
 \@ifx{#1\undefined}
}%
\providecommand \@ifnum [1]{%
 \ifnum #1\expandafter \@firstoftwo
 \else \expandafter \@secondoftwo
 \fi
}%
\providecommand \@ifx [1]{%
 \ifx #1\expandafter \@firstoftwo
 \else \expandafter \@secondoftwo
 \fi
}%
\providecommand \natexlab [1]{#1}%
\providecommand \enquote  [1]{``#1''}%
\providecommand \bibnamefont  [1]{#1}%
\providecommand \bibfnamefont [1]{#1}%
\providecommand \citenamefont [1]{#1}%
\providecommand \href@noop [0]{\@secondoftwo}%
\providecommand \href [0]{\begingroup \@sanitize@url \@href}%
\providecommand \@href[1]{\@@startlink{#1}\@@href}%
\providecommand \@@href[1]{\endgroup#1\@@endlink}%
\providecommand \@sanitize@url [0]{\catcode `\\12\catcode `\$12\catcode
  `\&12\catcode `\#12\catcode `\^12\catcode `\_12\catcode `\%12\relax}%
\providecommand \@@startlink[1]{}%
\providecommand \@@endlink[0]{}%
\providecommand \url  [0]{\begingroup\@sanitize@url \@url }%
\providecommand \@url [1]{\endgroup\@href {#1}{\urlprefix }}%
\providecommand \urlprefix  [0]{URL }%
\providecommand \Eprint [0]{\href }%
\providecommand \doibase [0]{https://doi.org/}%
\providecommand \selectlanguage [0]{\@gobble}%
\providecommand \bibinfo  [0]{\@secondoftwo}%
\providecommand \bibfield  [0]{\@secondoftwo}%
\providecommand \translation [1]{[#1]}%
\providecommand \BibitemOpen [0]{}%
\providecommand \bibitemStop [0]{}%
\providecommand \bibitemNoStop [0]{.\EOS\space}%
\providecommand \EOS [0]{\spacefactor3000\relax}%
\providecommand \BibitemShut  [1]{\csname bibitem#1\endcsname}%
\let\auto@bib@innerbib\@empty
\bibitem [{\citenamefont {Chen}\ \emph {et~al.}(2024)\citenamefont {Chen} \emph
  {et~al.}}]{Chen:2024aom}%
  \BibitemOpen
  \bibfield  {author} {\bibinfo {author} {\bibfnamefont {J.}~\bibnamefont
  {Chen}} \emph {et~al.},\ }\bibfield  {title} {\bibinfo {title} {{Properties
  of the QCD matter: review of selected results from the relativistic heavy ion
  collider beam energy scan (RHIC BES) program}},\ }\href
  {https://doi.org/10.1007/s41365-024-01591-2} {\bibfield  {journal} {\bibinfo
  {journal} {Nucl. Sci. Tech.}\ }\textbf {\bibinfo {volume} {35}},\ \bibinfo
  {pages} {214} (\bibinfo {year} {2024})},\ \Eprint
  {https://arxiv.org/abs/2407.02935} {arXiv:2407.02935 [nucl-ex]} \BibitemShut
  {NoStop}%
\bibitem [{\citenamefont {Mondal}(2024)}]{Mondal:2024sil}%
  \BibitemOpen
  \bibfield  {author} {\bibinfo {author} {\bibfnamefont {B.}~\bibnamefont
  {Mondal}} (\bibinfo {collaboration} {STAR}),\ }\bibfield  {title} {\bibinfo
  {title} {{Search for QCD Critical Point: Recent Results from STAR BES-I
  Program and Status of BES-II}},\ }\href@noop {} {\bibfield  {journal}
  {\bibinfo  {journal} {DAE Symp. Nucl. Phys.}\ }\textbf {\bibinfo {volume}
  {67}},\ \bibinfo {pages} {983} (\bibinfo {year} {2024})}\BibitemShut
  {NoStop}%
\bibitem [{Note1()}]{Note1}%
  \BibitemOpen
  \bibinfo {note} {STAR has recently reported a significant deviation of the
  kurtosis of the net-proton distribution from the noncritical baseline at
  $\protect \sqrt {s} = 19.6$\protect \,GeV \cite {STAR:2025zdq}. This
  corresponds to $(T,\mu _B) \approx (156,200)$\protect \,MeV at freeze-out and
  is therefore well within the range where a CEP has been excluded both from
  functional and lattice QCD, see, e.g., Refs.\ \cite {Fu:2019hdw, Gao:2020fbl,
  Gunkel:2021oya, Bazavov:2017dus, Borsanyi:2025dyp}.}\BibitemShut {Stop}%
\bibitem [{\citenamefont {Abdallah}\ \emph {et~al.}(2023)\citenamefont
  {Abdallah} \emph {et~al.}}]{STAR:2022etb}%
  \BibitemOpen
  \bibfield  {author} {\bibinfo {author} {\bibfnamefont {M.}~\bibnamefont
  {Abdallah}} \emph {et~al.} (\bibinfo {collaboration} {STAR}),\ }\bibfield
  {title} {\bibinfo {title} {{Higher-order cumulants and correlation functions
  of proton multiplicity distributions in sNN=3~GeV~Au+Au collisions at the
  RHIC STAR experiment}},\ }\href {https://doi.org/10.1103/PhysRevC.107.024908}
  {\bibfield  {journal} {\bibinfo  {journal} {Phys. Rev. C}\ }\textbf {\bibinfo
  {volume} {107}},\ \bibinfo {pages} {024908} (\bibinfo {year} {2023})},\
  \Eprint {https://arxiv.org/abs/2209.11940} {arXiv:2209.11940 [nucl-ex]}
  \BibitemShut {NoStop}%
\bibitem [{\citenamefont {Ablyazimov}\ \emph {et~al.}(2017)\citenamefont
  {Ablyazimov} \emph {et~al.}}]{CBM:2016kpk}%
  \BibitemOpen
  \bibfield  {author} {\bibinfo {author} {\bibfnamefont {T.}~\bibnamefont
  {Ablyazimov}} \emph {et~al.} (\bibinfo {collaboration} {CBM}),\ }\bibfield
  {title} {\bibinfo {title} {{Challenges in QCD matter physics --The scientific
  programme of the Compressed Baryonic Matter experiment at FAIR}},\ }\href
  {https://doi.org/10.1140/epja/i2017-12248-y} {\bibfield  {journal} {\bibinfo
  {journal} {Eur. Phys. J. A}\ }\textbf {\bibinfo {volume} {53}},\ \bibinfo
  {pages} {60} (\bibinfo {year} {2017})},\ \Eprint
  {https://arxiv.org/abs/1607.01487} {arXiv:1607.01487 [nucl-ex]} \BibitemShut
  {NoStop}%
\bibitem [{\citenamefont
  {Ma\'ckowiak-Paw\l{}owska}(2017)}]{Mackowiak-Pawlowska:2016qon}%
  \BibitemOpen
  \bibfield  {author} {\bibinfo {author} {\bibfnamefont {M.}~\bibnamefont
  {Ma\'ckowiak-Paw\l{}owska}} (\bibinfo {collaboration} {NA61/SHINE}),\
  }\bibfield  {title} {\bibinfo {title} {{Higher order moments of net-charge
  and multiplicity distributions in $p+p$ interactions at SPS energies from
  NA61/SHINE}},\ }\href {https://doi.org/10.5506/APhysPolBSupp.10.657}
  {\bibfield  {journal} {\bibinfo  {journal} {Acta Phys. Polon. Supp.}\
  }\textbf {\bibinfo {volume} {10}},\ \bibinfo {pages} {657} (\bibinfo {year}
  {2017})},\ \Eprint {https://arxiv.org/abs/1610.03838} {arXiv:1610.03838
  [nucl-ex]} \BibitemShut {NoStop}%
\bibitem [{\citenamefont {Kapishin}(2020)}]{Kapishin:2020cwk}%
  \BibitemOpen
  \bibfield  {author} {\bibinfo {author} {\bibfnamefont {M.}~\bibnamefont
  {Kapishin}},\ }\bibfield  {title} {\bibinfo {title} {{Heavy Ion BM@N and MPD
  Experiments at NICA}},\ }\href {https://doi.org/10.7566/JPSCP.32.010093}
  {\bibfield  {journal} {\bibinfo  {journal} {JPS Conf. Proc.}\ }\textbf
  {\bibinfo {volume} {32}},\ \bibinfo {pages} {010093} (\bibinfo {year}
  {2020})}\BibitemShut {NoStop}%
\bibitem [{\citenamefont {Sako}(2019)}]{Sako:2019hzh}%
  \BibitemOpen
  \bibfield  {author} {\bibinfo {author} {\bibfnamefont {H.}~\bibnamefont
  {Sako}} (\bibinfo {collaboration} {J-PARC-HI}),\ }\bibfield  {title}
  {\bibinfo {title} {{Studies of extremely dense matter in heavy-ion collisions
  at J-PARC}},\ }\href {https://doi.org/10.1016/j.nuclphysa.2018.11.027}
  {\bibfield  {journal} {\bibinfo  {journal} {Nucl. Phys. A}\ }\textbf
  {\bibinfo {volume} {982}},\ \bibinfo {pages} {959} (\bibinfo {year}
  {2019})}\BibitemShut {NoStop}%
\bibitem [{\citenamefont {Bellwied}\ \emph {et~al.}(2015)\citenamefont
  {Bellwied}, \citenamefont {Borsanyi}, \citenamefont {Fodor}, \citenamefont
  {Guenther}, \citenamefont {Katz}, \citenamefont {Ratti},\ and\ \citenamefont
  {Szabo}}]{Bellwied:2015rza}%
  \BibitemOpen
  \bibfield  {author} {\bibinfo {author} {\bibfnamefont {R.}~\bibnamefont
  {Bellwied}}, \bibinfo {author} {\bibfnamefont {S.}~\bibnamefont {Borsanyi}},
  \bibinfo {author} {\bibfnamefont {Z.}~\bibnamefont {Fodor}}, \bibinfo
  {author} {\bibfnamefont {J.}~\bibnamefont {Guenther}}, \bibinfo {author}
  {\bibfnamefont {S.~D.}\ \bibnamefont {Katz}}, \bibinfo {author}
  {\bibfnamefont {C.}~\bibnamefont {Ratti}},\ and\ \bibinfo {author}
  {\bibfnamefont {K.~K.}\ \bibnamefont {Szabo}},\ }\bibfield  {title} {\bibinfo
  {title} {{The QCD phase diagram from analytic continuation}},\ }\href
  {https://doi.org/10.1016/j.physletb.2015.11.011} {\bibfield  {journal}
  {\bibinfo  {journal} {Phys. Lett.}\ }\textbf {\bibinfo {volume} {B751}},\
  \bibinfo {pages} {559} (\bibinfo {year} {2015})},\ \Eprint
  {https://arxiv.org/abs/1507.07510} {arXiv:1507.07510 [hep-lat]} \BibitemShut
  {NoStop}%
\bibitem [{\citenamefont {Bazavov}\ \emph
  {et~al.}(2019{\natexlab{a}})\citenamefont {Bazavov} \emph
  {et~al.}}]{Bazavov:2018mes}%
  \BibitemOpen
  \bibfield  {author} {\bibinfo {author} {\bibfnamefont {A.}~\bibnamefont
  {Bazavov}} \emph {et~al.} (\bibinfo {collaboration} {HotQCD}),\ }\bibfield
  {title} {\bibinfo {title} {{Chiral crossover in QCD at zero and non-zero
  chemical potentials}},\ }\href
  {https://doi.org/10.1016/j.physletb.2019.05.013} {\bibfield  {journal}
  {\bibinfo  {journal} {Phys. Lett.}\ }\textbf {\bibinfo {volume} {B795}},\
  \bibinfo {pages} {15} (\bibinfo {year} {2019}{\natexlab{a}})},\ \Eprint
  {https://arxiv.org/abs/1812.08235} {arXiv:1812.08235 [hep-lat]} \BibitemShut
  {NoStop}%
\bibitem [{\citenamefont {Fu}\ \emph {et~al.}(2020)\citenamefont {Fu},
  \citenamefont {Pawlowski},\ and\ \citenamefont {Rennecke}}]{Fu:2019hdw}%
  \BibitemOpen
  \bibfield  {author} {\bibinfo {author} {\bibfnamefont {W.-j.}\ \bibnamefont
  {Fu}}, \bibinfo {author} {\bibfnamefont {J.~M.}\ \bibnamefont {Pawlowski}},\
  and\ \bibinfo {author} {\bibfnamefont {F.}~\bibnamefont {Rennecke}},\
  }\bibfield  {title} {\bibinfo {title} {{QCD phase structure at finite
  temperature and density}},\ }\href
  {https://doi.org/10.1103/PhysRevD.101.054032} {\bibfield  {journal} {\bibinfo
   {journal} {Phys. Rev. D}\ }\textbf {\bibinfo {volume} {101}},\ \bibinfo
  {pages} {054032} (\bibinfo {year} {2020})},\ \Eprint
  {https://arxiv.org/abs/1909.02991} {arXiv:1909.02991 [hep-ph]} \BibitemShut
  {NoStop}%
\bibitem [{\citenamefont {Gao}\ and\ \citenamefont
  {Pawlowski}(2021)}]{Gao:2020fbl}%
  \BibitemOpen
  \bibfield  {author} {\bibinfo {author} {\bibfnamefont {F.}~\bibnamefont
  {Gao}}\ and\ \bibinfo {author} {\bibfnamefont {J.~M.}\ \bibnamefont
  {Pawlowski}},\ }\bibfield  {title} {\bibinfo {title} {{Chiral phase structure
  and critical end point in QCD}},\ }\href
  {https://doi.org/10.1016/j.physletb.2021.136584} {\bibfield  {journal}
  {\bibinfo  {journal} {Phys. Lett. B}\ }\textbf {\bibinfo {volume} {820}},\
  \bibinfo {pages} {136584} (\bibinfo {year} {2021})},\ \Eprint
  {https://arxiv.org/abs/2010.13705} {arXiv:2010.13705 [hep-ph]} \BibitemShut
  {NoStop}%
\bibitem [{\citenamefont {Gunkel}\ and\ \citenamefont
  {Fischer}(2021)}]{Gunkel:2021oya}%
  \BibitemOpen
  \bibfield  {author} {\bibinfo {author} {\bibfnamefont {P.~J.}\ \bibnamefont
  {Gunkel}}\ and\ \bibinfo {author} {\bibfnamefont {C.~S.}\ \bibnamefont
  {Fischer}},\ }\href@noop {} {\bibinfo {title} {{Locating the critical
  endpoint of QCD: mesonic backcoupling effects}}} (\bibinfo {year} {2021}),\
  \Eprint {https://arxiv.org/abs/2106.08356} {arXiv:2106.08356 [hep-ph]}
  \BibitemShut {NoStop}%
\bibitem [{\citenamefont {Pisarski}\ and\ \citenamefont
  {Rennecke}(2021)}]{Pisarski:2021qof}%
  \BibitemOpen
  \bibfield  {author} {\bibinfo {author} {\bibfnamefont {R.~D.}\ \bibnamefont
  {Pisarski}}\ and\ \bibinfo {author} {\bibfnamefont {F.}~\bibnamefont
  {Rennecke}},\ }\bibfield  {title} {\bibinfo {title} {{Signatures of Moat
  Regimes in Heavy-Ion Collisions}},\ }\href
  {https://doi.org/10.1103/PhysRevLett.127.152302} {\bibfield  {journal}
  {\bibinfo  {journal} {Phys. Rev. Lett.}\ }\textbf {\bibinfo {volume} {127}},\
  \bibinfo {pages} {152302} (\bibinfo {year} {2021})},\ \Eprint
  {https://arxiv.org/abs/2103.06890} {arXiv:2103.06890 [hep-ph]} \BibitemShut
  {NoStop}%
\bibitem [{\citenamefont {Fu}\ \emph {et~al.}(2025)\citenamefont {Fu},
  \citenamefont {Pawlowski}, \citenamefont {Pisarski}, \citenamefont
  {Rennecke}, \citenamefont {Wen},\ and\ \citenamefont {Yin}}]{Fu:2024rto}%
  \BibitemOpen
  \bibfield  {author} {\bibinfo {author} {\bibfnamefont {W.-j.}\ \bibnamefont
  {Fu}}, \bibinfo {author} {\bibfnamefont {J.~M.}\ \bibnamefont {Pawlowski}},
  \bibinfo {author} {\bibfnamefont {R.~D.}\ \bibnamefont {Pisarski}}, \bibinfo
  {author} {\bibfnamefont {F.}~\bibnamefont {Rennecke}}, \bibinfo {author}
  {\bibfnamefont {R.}~\bibnamefont {Wen}},\ and\ \bibinfo {author}
  {\bibfnamefont {S.}~\bibnamefont {Yin}},\ }\bibfield  {title} {\bibinfo
  {title} {{QCD moat regime and its real-time properties}},\ }\href
  {https://doi.org/10.1103/PhysRevD.111.094026} {\bibfield  {journal} {\bibinfo
   {journal} {Phys. Rev. D}\ }\textbf {\bibinfo {volume} {111}},\ \bibinfo
  {pages} {094026} (\bibinfo {year} {2025})},\ \Eprint
  {https://arxiv.org/abs/2412.15949} {arXiv:2412.15949 [hep-ph]} \BibitemShut
  {NoStop}%
\bibitem [{\citenamefont {Motta}\ \emph {et~al.}(2023)\citenamefont {Motta},
  \citenamefont {Bernhardt}, \citenamefont {Buballa},\ and\ \citenamefont
  {Fischer}}]{Motta:2023pks}%
  \BibitemOpen
  \bibfield  {author} {\bibinfo {author} {\bibfnamefont {T.~F.}\ \bibnamefont
  {Motta}}, \bibinfo {author} {\bibfnamefont {J.}~\bibnamefont {Bernhardt}},
  \bibinfo {author} {\bibfnamefont {M.}~\bibnamefont {Buballa}},\ and\ \bibinfo
  {author} {\bibfnamefont {C.~S.}\ \bibnamefont {Fischer}},\ }\bibfield
  {title} {\bibinfo {title} {{Toward a stability analysis of inhomogeneous
  phases in QCD}},\ }\href {https://doi.org/10.1103/PhysRevD.108.114019}
  {\bibfield  {journal} {\bibinfo  {journal} {Phys. Rev. D}\ }\textbf {\bibinfo
  {volume} {108}},\ \bibinfo {pages} {114019} (\bibinfo {year} {2023})},\
  \Eprint {https://arxiv.org/abs/2306.09749} {arXiv:2306.09749 [hep-ph]}
  \BibitemShut {NoStop}%
\bibitem [{\citenamefont {Motta}\ \emph {et~al.}(2024)\citenamefont {Motta},
  \citenamefont {Bernhardt}, \citenamefont {Buballa},\ and\ \citenamefont
  {Fischer}}]{Motta:2024agi}%
  \BibitemOpen
  \bibfield  {author} {\bibinfo {author} {\bibfnamefont {T.~F.}\ \bibnamefont
  {Motta}}, \bibinfo {author} {\bibfnamefont {J.}~\bibnamefont {Bernhardt}},
  \bibinfo {author} {\bibfnamefont {M.}~\bibnamefont {Buballa}},\ and\ \bibinfo
  {author} {\bibfnamefont {C.~S.}\ \bibnamefont {Fischer}},\ }\bibfield
  {title} {\bibinfo {title} {{Inhomogenuous instabilities at large chemical
  potential in a rainbow-ladder QCD model}},\ }\href
  {https://doi.org/10.1103/PhysRevD.110.074014} {\bibfield  {journal} {\bibinfo
   {journal} {Phys. Rev. D}\ }\textbf {\bibinfo {volume} {110}},\ \bibinfo
  {pages} {074014} (\bibinfo {year} {2024})},\ \Eprint
  {https://arxiv.org/abs/2406.00205} {arXiv:2406.00205 [hep-ph]} \BibitemShut
  {NoStop}%
\bibitem [{\citenamefont {Motta}\ \emph {et~al.}(2025)\citenamefont {Motta},
  \citenamefont {Bernhardt}, \citenamefont {Buballa},\ and\ \citenamefont
  {Fischer}}]{Motta:2024rvk}%
  \BibitemOpen
  \bibfield  {author} {\bibinfo {author} {\bibfnamefont {T.~F.}\ \bibnamefont
  {Motta}}, \bibinfo {author} {\bibfnamefont {J.}~\bibnamefont {Bernhardt}},
  \bibinfo {author} {\bibfnamefont {M.}~\bibnamefont {Buballa}},\ and\ \bibinfo
  {author} {\bibfnamefont {C.~S.}\ \bibnamefont {Fischer}},\ }\bibfield
  {title} {\bibinfo {title} {{New tool to detect inhomogeneous chiral-symmetry
  breaking}},\ }\href {https://doi.org/10.1103/PhysRevD.111.074030} {\bibfield
  {journal} {\bibinfo  {journal} {Phys. Rev. D}\ }\textbf {\bibinfo {volume}
  {111}},\ \bibinfo {pages} {074030} (\bibinfo {year} {2025})},\ \Eprint
  {https://arxiv.org/abs/2411.02285} {arXiv:2411.02285 [hep-ph]} \BibitemShut
  {NoStop}%
\bibitem [{\citenamefont {Fulde}\ and\ \citenamefont
  {Ferrell}(1964)}]{Fulde:1964zz}%
  \BibitemOpen
  \bibfield  {author} {\bibinfo {author} {\bibfnamefont {P.}~\bibnamefont
  {Fulde}}\ and\ \bibinfo {author} {\bibfnamefont {R.~A.}\ \bibnamefont
  {Ferrell}},\ }\bibfield  {title} {\bibinfo {title} {{Superconductivity in a
  Strong Spin-Exchange Field}},\ }\href
  {https://doi.org/10.1103/PhysRev.135.A550} {\bibfield  {journal} {\bibinfo
  {journal} {Phys. Rev.}\ }\textbf {\bibinfo {volume} {135}},\ \bibinfo {pages}
  {A550} (\bibinfo {year} {1964})}\BibitemShut {NoStop}%
\bibitem [{\citenamefont {Hornreich}\ \emph {et~al.}(1975)\citenamefont
  {Hornreich}, \citenamefont {Luban},\ and\ \citenamefont
  {Shtrikman}}]{Hornreich:1975zz}%
  \BibitemOpen
  \bibfield  {author} {\bibinfo {author} {\bibfnamefont {R.~M.}\ \bibnamefont
  {Hornreich}}, \bibinfo {author} {\bibfnamefont {M.}~\bibnamefont {Luban}},\
  and\ \bibinfo {author} {\bibfnamefont {S.}~\bibnamefont {Shtrikman}},\
  }\bibfield  {title} {\bibinfo {title} {{Critical Behavior at the Onset of
  $\vec{k}$--Space Instability on the lamda Line}},\ }\href
  {https://doi.org/10.1103/PhysRevLett.35.1678} {\bibfield  {journal} {\bibinfo
   {journal} {Phys. Rev. Lett.}\ }\textbf {\bibinfo {volume} {35}},\ \bibinfo
  {pages} {1678} (\bibinfo {year} {1975})}\BibitemShut {NoStop}%
\bibitem [{\citenamefont {Seul}\ and\ \citenamefont
  {Andelman}(1995)}]{Seul-Andelman:1995}%
  \BibitemOpen
  \bibfield  {author} {\bibinfo {author} {\bibfnamefont {M.}~\bibnamefont
  {Seul}}\ and\ \bibinfo {author} {\bibfnamefont {D.}~\bibnamefont
  {Andelman}},\ }\bibfield  {title} {\bibinfo {title} {Domain shapes and
  patterns: The phenomenology of modulated phases},\ }\href
  {https://doi.org/10.1126/science.267.5197.476} {\bibfield  {journal}
  {\bibinfo  {journal} {Science}\ }\textbf {\bibinfo {volume} {267}},\ \bibinfo
  {pages} {476} (\bibinfo {year} {1995})}\BibitemShut {NoStop}%
\bibitem [{\citenamefont {Chakrabarty}\ and\ \citenamefont
  {Nussinov}(2011)}]{Chakrabarty_2011}%
  \BibitemOpen
  \bibfield  {author} {\bibinfo {author} {\bibfnamefont {S.}~\bibnamefont
  {Chakrabarty}}\ and\ \bibinfo {author} {\bibfnamefont {Z.}~\bibnamefont
  {Nussinov}},\ }\bibfield  {title} {\bibinfo {title} {Modulation and
  correlation lengths in systems with competing interactions},\ }\bibfield
  {journal} {\bibinfo  {journal} {Physical Review B}\ }\textbf {\bibinfo
  {volume} {84}},\ \href {https://doi.org/10.1103/physrevb.84.144402}
  {10.1103/physrevb.84.144402} (\bibinfo {year} {2011})\BibitemShut {NoStop}%
\bibitem [{\citenamefont {Sedrakyan}\ \emph {et~al.}(2014)\citenamefont
  {Sedrakyan}, \citenamefont {Glazman},\ and\ \citenamefont
  {Kamenev}}]{Sedrakyan:2013qja}%
  \BibitemOpen
  \bibfield  {author} {\bibinfo {author} {\bibfnamefont {T.~A.}\ \bibnamefont
  {Sedrakyan}}, \bibinfo {author} {\bibfnamefont {L.~I.}\ \bibnamefont
  {Glazman}},\ and\ \bibinfo {author} {\bibfnamefont {A.}~\bibnamefont
  {Kamenev}},\ }\bibfield  {title} {\bibinfo {title} {{Absence of Bose
  condensation on lattices with moat bands}},\ }\href
  {https://doi.org/10.1103/PhysRevB.89.201112} {\bibfield  {journal} {\bibinfo
  {journal} {Phys. Rev. B}\ }\textbf {\bibinfo {volume} {89}},\ \bibinfo
  {pages} {201112} (\bibinfo {year} {2014})},\ \Eprint
  {https://arxiv.org/abs/1303.7272} {arXiv:1303.7272 [cond-mat.quant-gas]}
  \BibitemShut {NoStop}%
\bibitem [{\citenamefont {Buballa}\ and\ \citenamefont
  {Carignano}(2015)}]{Buballa:2014tba}%
  \BibitemOpen
  \bibfield  {author} {\bibinfo {author} {\bibfnamefont {M.}~\bibnamefont
  {Buballa}}\ and\ \bibinfo {author} {\bibfnamefont {S.}~\bibnamefont
  {Carignano}},\ }\bibfield  {title} {\bibinfo {title} {{Inhomogeneous chiral
  condensates}},\ }\href {https://doi.org/10.1016/j.ppnp.2014.11.001}
  {\bibfield  {journal} {\bibinfo  {journal} {Prog. Part. Nucl. Phys.}\
  }\textbf {\bibinfo {volume} {81}},\ \bibinfo {pages} {39} (\bibinfo {year}
  {2015})},\ \Eprint {https://arxiv.org/abs/1406.1367} {arXiv:1406.1367
  [hep-ph]} \BibitemShut {NoStop}%
\bibitem [{\citenamefont {Schindler}\ \emph {et~al.}(2020)\citenamefont
  {Schindler}, \citenamefont {Schindler}, \citenamefont {Medina},\ and\
  \citenamefont {Ogilvie}}]{Schindler:2019ugo}%
  \BibitemOpen
  \bibfield  {author} {\bibinfo {author} {\bibfnamefont {M.~A.}\ \bibnamefont
  {Schindler}}, \bibinfo {author} {\bibfnamefont {S.~T.}\ \bibnamefont
  {Schindler}}, \bibinfo {author} {\bibfnamefont {L.}~\bibnamefont {Medina}},\
  and\ \bibinfo {author} {\bibfnamefont {M.~C.}\ \bibnamefont {Ogilvie}},\
  }\bibfield  {title} {\bibinfo {title} {{Universality of Pattern Formation}},\
  }\href {https://doi.org/10.1103/PhysRevD.102.114510} {\bibfield  {journal}
  {\bibinfo  {journal} {Phys. Rev. D}\ }\textbf {\bibinfo {volume} {102}},\
  \bibinfo {pages} {114510} (\bibinfo {year} {2020})},\ \Eprint
  {https://arxiv.org/abs/1906.07288} {arXiv:1906.07288 [hep-lat]} \BibitemShut
  {NoStop}%
\bibitem [{\citenamefont {Pisarski}\ \emph {et~al.}(2020)\citenamefont
  {Pisarski}, \citenamefont {Tsvelik},\ and\ \citenamefont
  {Valgushev}}]{Pisarski:2020dnx}%
  \BibitemOpen
  \bibfield  {author} {\bibinfo {author} {\bibfnamefont {R.~D.}\ \bibnamefont
  {Pisarski}}, \bibinfo {author} {\bibfnamefont {A.~M.}\ \bibnamefont
  {Tsvelik}},\ and\ \bibinfo {author} {\bibfnamefont {S.}~\bibnamefont
  {Valgushev}},\ }\bibfield  {title} {\bibinfo {title} {{How transverse thermal
  fluctuations disorder a condensate of chiral spirals into a quantum spin
  liquid}},\ }\href {https://doi.org/10.1103/PhysRevD.102.016015} {\bibfield
  {journal} {\bibinfo  {journal} {Phys. Rev. D}\ }\textbf {\bibinfo {volume}
  {102}},\ \bibinfo {pages} {016015} (\bibinfo {year} {2020})},\ \Eprint
  {https://arxiv.org/abs/2005.10259} {arXiv:2005.10259 [hep-ph]} \BibitemShut
  {NoStop}%
\bibitem [{\citenamefont {Schindler}\ \emph {et~al.}(2021)\citenamefont
  {Schindler}, \citenamefont {Schindler},\ and\ \citenamefont
  {Ogilvie}}]{Schindler:2021otf}%
  \BibitemOpen
  \bibfield  {author} {\bibinfo {author} {\bibfnamefont {M.~A.}\ \bibnamefont
  {Schindler}}, \bibinfo {author} {\bibfnamefont {S.~T.}\ \bibnamefont
  {Schindler}},\ and\ \bibinfo {author} {\bibfnamefont {M.~C.}\ \bibnamefont
  {Ogilvie}},\ }\bibfield  {title} {\bibinfo {title} {{$\mathcal PT$ symmetry,
  pattern formation, and finite-density QCD}}\ }\href
  {https://doi.org/10.1088/1742-6596/2038/1/012022}
  {10.1088/1742-6596/2038/1/012022} (\bibinfo {year} {2021}),\ \Eprint
  {https://arxiv.org/abs/2106.07092} {arXiv:2106.07092 [hep-lat]} \BibitemShut
  {NoStop}%
\bibitem [{\citenamefont {Nussinov}\ \emph {et~al.}(2025)\citenamefont
  {Nussinov}, \citenamefont {Ogilvie}, \citenamefont {Pannullo}, \citenamefont
  {Pisarski}, \citenamefont {Rennecke}, \citenamefont {Schindler},\ and\
  \citenamefont {Winstel}}]{Nussinov:2024erh}%
  \BibitemOpen
  \bibfield  {author} {\bibinfo {author} {\bibfnamefont {Z.}~\bibnamefont
  {Nussinov}}, \bibinfo {author} {\bibfnamefont {M.~C.}\ \bibnamefont
  {Ogilvie}}, \bibinfo {author} {\bibfnamefont {L.}~\bibnamefont {Pannullo}},
  \bibinfo {author} {\bibfnamefont {R.~D.}\ \bibnamefont {Pisarski}}, \bibinfo
  {author} {\bibfnamefont {F.}~\bibnamefont {Rennecke}}, \bibinfo {author}
  {\bibfnamefont {S.~T.}\ \bibnamefont {Schindler}},\ and\ \bibinfo {author}
  {\bibfnamefont {M.}~\bibnamefont {Winstel}},\ }\bibfield  {title} {\bibinfo
  {title} {{Dilepton Production from Moaton Quasiparticles}},\ }\href
  {https://doi.org/10.1103/1f8w-p4m4} {\bibfield  {journal} {\bibinfo
  {journal} {Phys. Rev. Lett.}\ }\textbf {\bibinfo {volume} {135}},\ \bibinfo
  {pages} {101904} (\bibinfo {year} {2025})},\ \Eprint
  {https://arxiv.org/abs/2410.22418} {arXiv:2410.22418 [hep-ph]} \BibitemShut
  {NoStop}%
\bibitem [{\citenamefont {Pisarski}\ \emph {et~al.}(2021)\citenamefont
  {Pisarski}, \citenamefont {Rennecke}, \citenamefont {Tsvelik},\ and\
  \citenamefont {Valgushev}}]{Pisarski:2020gkx}%
  \BibitemOpen
  \bibfield  {author} {\bibinfo {author} {\bibfnamefont {R.~D.}\ \bibnamefont
  {Pisarski}}, \bibinfo {author} {\bibfnamefont {F.}~\bibnamefont {Rennecke}},
  \bibinfo {author} {\bibfnamefont {A.}~\bibnamefont {Tsvelik}},\ and\ \bibinfo
  {author} {\bibfnamefont {S.}~\bibnamefont {Valgushev}},\ }\bibfield  {title}
  {\bibinfo {title} {{The Lifshitz Regime and its Experimental Signals}},\
  }\href {https://doi.org/10.1016/j.nuclphysa.2020.121910} {\bibfield
  {journal} {\bibinfo  {journal} {Nucl. Phys. A}\ }\textbf {\bibinfo {volume}
  {1005}},\ \bibinfo {pages} {121910} (\bibinfo {year} {2021})},\ \Eprint
  {https://arxiv.org/abs/2005.00045} {arXiv:2005.00045 [nucl-th]} \BibitemShut
  {NoStop}%
\bibitem [{\citenamefont {Rennecke}\ and\ \citenamefont
  {Pisarski}(2022)}]{Rennecke:2021ovl}%
  \BibitemOpen
  \bibfield  {author} {\bibinfo {author} {\bibfnamefont {F.}~\bibnamefont
  {Rennecke}}\ and\ \bibinfo {author} {\bibfnamefont {R.~D.}\ \bibnamefont
  {Pisarski}},\ }\bibfield  {title} {\bibinfo {title} {{Moat Regimes in QCD and
  their Signatures in Heavy-Ion Collisions}},\ }\href
  {https://doi.org/10.22323/1.400.0016} {\bibfield  {journal} {\bibinfo
  {journal} {PoS}\ }\textbf {\bibinfo {volume} {CPOD2021}},\ \bibinfo {pages}
  {016} (\bibinfo {year} {2022})},\ \Eprint {https://arxiv.org/abs/2110.02625}
  {arXiv:2110.02625 [hep-ph]} \BibitemShut {NoStop}%
\bibitem [{\citenamefont {Rennecke}\ \emph {et~al.}(2023)\citenamefont
  {Rennecke}, \citenamefont {Pisarski},\ and\ \citenamefont
  {Rischke}}]{Rennecke:2023xhc}%
  \BibitemOpen
  \bibfield  {author} {\bibinfo {author} {\bibfnamefont {F.}~\bibnamefont
  {Rennecke}}, \bibinfo {author} {\bibfnamefont {R.~D.}\ \bibnamefont
  {Pisarski}},\ and\ \bibinfo {author} {\bibfnamefont {D.~H.}\ \bibnamefont
  {Rischke}},\ }\href@noop {} {\bibinfo {title} {{Particle Interferometry in a
  Moat Regime}}} (\bibinfo {year} {2023}),\ \Eprint
  {https://arxiv.org/abs/2301.11484} {arXiv:2301.11484 [hep-ph]} \BibitemShut
  {NoStop}%
\bibitem [{\citenamefont {Koenigstein}\ \emph {et~al.}(2022)\citenamefont
  {Koenigstein}, \citenamefont {Pannullo}, \citenamefont {Rechenberger},
  \citenamefont {Steil},\ and\ \citenamefont {Winstel}}]{Koenigstein:2021llr}%
  \BibitemOpen
  \bibfield  {author} {\bibinfo {author} {\bibfnamefont {A.}~\bibnamefont
  {Koenigstein}}, \bibinfo {author} {\bibfnamefont {L.}~\bibnamefont
  {Pannullo}}, \bibinfo {author} {\bibfnamefont {S.}~\bibnamefont
  {Rechenberger}}, \bibinfo {author} {\bibfnamefont {M.~J.}\ \bibnamefont
  {Steil}},\ and\ \bibinfo {author} {\bibfnamefont {M.}~\bibnamefont
  {Winstel}},\ }\bibfield  {title} {\bibinfo {title} {{Detecting inhomogeneous
  chiral condensation from the bosonic two-point function in the (1 +
  1)-dimensional Gross\textendash{}Neveu model in the mean-field
  approximation*}},\ }\href {https://doi.org/10.1088/1751-8121/ac820a}
  {\bibfield  {journal} {\bibinfo  {journal} {J. Phys. A}\ }\textbf {\bibinfo
  {volume} {55}},\ \bibinfo {pages} {375402} (\bibinfo {year} {2022})},\
  \Eprint {https://arxiv.org/abs/2112.07024} {arXiv:2112.07024 [hep-ph]}
  \BibitemShut {NoStop}%
\bibitem [{\citenamefont {Pannullo}\ and\ \citenamefont
  {Winstel}(2023)}]{Pannullo:2023one}%
  \BibitemOpen
  \bibfield  {author} {\bibinfo {author} {\bibfnamefont {L.}~\bibnamefont
  {Pannullo}}\ and\ \bibinfo {author} {\bibfnamefont {M.}~\bibnamefont
  {Winstel}},\ }\bibfield  {title} {\bibinfo {title} {{Absence of inhomogeneous
  chiral phases in (2+1)-dimensional four-fermion and Yukawa models}},\ }\href
  {https://doi.org/10.1103/PhysRevD.108.036011} {\bibfield  {journal} {\bibinfo
   {journal} {Phys. Rev. D}\ }\textbf {\bibinfo {volume} {108}},\ \bibinfo
  {pages} {036011} (\bibinfo {year} {2023})},\ \Eprint
  {https://arxiv.org/abs/2305.09444} {arXiv:2305.09444 [hep-ph]} \BibitemShut
  {NoStop}%
\bibitem [{\citenamefont {Pannullo}(2023)}]{Pannullo:2023cat}%
  \BibitemOpen
  \bibfield  {author} {\bibinfo {author} {\bibfnamefont {L.}~\bibnamefont
  {Pannullo}},\ }\bibfield  {title} {\bibinfo {title} {{Inhomogeneous
  condensation in the Gross-Neveu model in noninteger spatial dimensions
  1\ensuremath{\leq}d\ensuremath{<}3}},\ }\href
  {https://doi.org/10.1103/PhysRevD.108.036022} {\bibfield  {journal} {\bibinfo
   {journal} {Phys. Rev. D}\ }\textbf {\bibinfo {volume} {108}},\ \bibinfo
  {pages} {036022} (\bibinfo {year} {2023})},\ \Eprint
  {https://arxiv.org/abs/2306.16290} {arXiv:2306.16290 [hep-ph]} \BibitemShut
  {NoStop}%
\bibitem [{\citenamefont {Koenigstein}\ and\ \citenamefont
  {Pannullo}(2024)}]{Koenigstein:2023yzv}%
  \BibitemOpen
  \bibfield  {author} {\bibinfo {author} {\bibfnamefont {A.}~\bibnamefont
  {Koenigstein}}\ and\ \bibinfo {author} {\bibfnamefont {L.}~\bibnamefont
  {Pannullo}},\ }\bibfield  {title} {\bibinfo {title} {{Inhomogeneous
  condensation in the Gross-Neveu model in noninteger spatial dimensions
  1{\ensuremath{\leq}}d{\ensuremath{<}}3. II. Nonzero temperature and chemical
  potential}},\ }\href {https://doi.org/10.1103/PhysRevD.109.056015} {\bibfield
   {journal} {\bibinfo  {journal} {Phys. Rev. D}\ }\textbf {\bibinfo {volume}
  {109}},\ \bibinfo {pages} {056015} (\bibinfo {year} {2024})},\ \Eprint
  {https://arxiv.org/abs/2312.04904} {arXiv:2312.04904 [hep-ph]} \BibitemShut
  {NoStop}%
\bibitem [{\citenamefont {Winstel}(2024)}]{Winstel:2024dqu}%
  \BibitemOpen
  \bibfield  {author} {\bibinfo {author} {\bibfnamefont {M.}~\bibnamefont
  {Winstel}},\ }\bibfield  {title} {\bibinfo {title} {{Spatially oscillating
  correlation functions in (2+1)-dimensional four-fermion models: The mixing of
  scalar and vector modes at finite density}},\ }\href
  {https://doi.org/10.1103/PhysRevD.110.034008} {\bibfield  {journal} {\bibinfo
   {journal} {Phys. Rev. D}\ }\textbf {\bibinfo {volume} {110}},\ \bibinfo
  {pages} {034008} (\bibinfo {year} {2024})},\ \Eprint
  {https://arxiv.org/abs/2403.07430} {arXiv:2403.07430 [hep-ph]} \BibitemShut
  {NoStop}%
\bibitem [{\citenamefont {Pannullo}\ \emph {et~al.}(2024)\citenamefont
  {Pannullo}, \citenamefont {Wagner},\ and\ \citenamefont
  {Winstel}}]{Pannullo:2024sov}%
  \BibitemOpen
  \bibfield  {author} {\bibinfo {author} {\bibfnamefont {L.}~\bibnamefont
  {Pannullo}}, \bibinfo {author} {\bibfnamefont {M.}~\bibnamefont {Wagner}},\
  and\ \bibinfo {author} {\bibfnamefont {M.}~\bibnamefont {Winstel}},\
  }\bibfield  {title} {\bibinfo {title} {{Regularization effects in the
  Nambu\textendash{}Jona-Lasinio model: Strong scheme dependence of
  inhomogeneous phases and persistence of the moat regime}},\ }\href
  {https://doi.org/10.1103/PhysRevD.110.076006} {\bibfield  {journal} {\bibinfo
   {journal} {Phys. Rev. D}\ }\textbf {\bibinfo {volume} {110}},\ \bibinfo
  {pages} {076006} (\bibinfo {year} {2024})},\ \Eprint
  {https://arxiv.org/abs/2406.11312} {arXiv:2406.11312 [hep-ph]} \BibitemShut
  {NoStop}%
\bibitem [{\citenamefont {T\"opfel}\ \emph {et~al.}(2024)\citenamefont
  {T\"opfel}, \citenamefont {Pawlowski},\ and\ \citenamefont
  {Braun}}]{Topfel:2024iop}%
  \BibitemOpen
  \bibfield  {author} {\bibinfo {author} {\bibfnamefont {S.}~\bibnamefont
  {T\"opfel}}, \bibinfo {author} {\bibfnamefont {J.~M.}\ \bibnamefont
  {Pawlowski}},\ and\ \bibinfo {author} {\bibfnamefont {J.}~\bibnamefont
  {Braun}},\ }\bibfield  {title} {\bibinfo {title} {Phase structure of quark
  matter and in-medium properties of mesons from callan-symanzik flows},\
  }\href@noop {} {\  (\bibinfo {year} {2024})},\ \Eprint
  {https://arxiv.org/abs/2412.16059} {arXiv:2412.16059 [hep-ph]} \BibitemShut
  {NoStop}%
\bibitem [{\citenamefont {Cao}(2025)}]{Cao:2025zvh}%
  \BibitemOpen
  \bibfield  {author} {\bibinfo {author} {\bibfnamefont {G.}~\bibnamefont
  {Cao}},\ }\bibfield  {title} {\bibinfo {title} {{The moat regimes within
  $2+1$ flavor Polyakov-Quark-Meson model}},\ }\href@noop {} {\  (\bibinfo
  {year} {2025})},\ \Eprint {https://arxiv.org/abs/2504.18874}
  {arXiv:2504.18874 [hep-ph]} \BibitemShut {NoStop}%
\bibitem [{\citenamefont {Motta}\ and\ \citenamefont
  {Krein}(2025)}]{Motta:2025xop}%
  \BibitemOpen
  \bibfield  {author} {\bibinfo {author} {\bibfnamefont {T.~F.}\ \bibnamefont
  {Motta}}\ and\ \bibinfo {author} {\bibfnamefont {G.}~\bibnamefont {Krein}},\
  }\bibfield  {title} {\bibinfo {title} {{The Quark Structure of the Nucleon
  and Moat Regimes in Nuclear Matter}},\ }\href@noop {} {\  (\bibinfo {year}
  {2025})},\ \Eprint {https://arxiv.org/abs/2508.09287} {arXiv:2508.09287
  [nucl-th]} \BibitemShut {NoStop}%
\bibitem [{\citenamefont {Partyka}\ and\ \citenamefont
  {Sadzikowski}(2009)}]{Partyka:2008sv}%
  \BibitemOpen
  \bibfield  {author} {\bibinfo {author} {\bibfnamefont {T.~L.}\ \bibnamefont
  {Partyka}}\ and\ \bibinfo {author} {\bibfnamefont {M.}~\bibnamefont
  {Sadzikowski}},\ }\bibfield  {title} {\bibinfo {title} {{Phase diagram of the
  non-uniform chiral condensate in different regularization schemes at T=0}},\
  }\href {https://doi.org/10.1088/0954-3899/36/2/025004} {\bibfield  {journal}
  {\bibinfo  {journal} {J. Phys. G}\ }\textbf {\bibinfo {volume} {36}},\
  \bibinfo {pages} {025004} (\bibinfo {year} {2009})},\ \Eprint
  {https://arxiv.org/abs/0811.4616} {arXiv:0811.4616 [hep-ph]} \BibitemShut
  {NoStop}%
\bibitem [{\citenamefont {Buballa}\ \emph {et~al.}(2021)\citenamefont
  {Buballa}, \citenamefont {Kurth}, \citenamefont {Wagner},\ and\ \citenamefont
  {Winstel}}]{Buballa:2020nsi}%
  \BibitemOpen
  \bibfield  {author} {\bibinfo {author} {\bibfnamefont {M.}~\bibnamefont
  {Buballa}}, \bibinfo {author} {\bibfnamefont {L.}~\bibnamefont {Kurth}},
  \bibinfo {author} {\bibfnamefont {M.}~\bibnamefont {Wagner}},\ and\ \bibinfo
  {author} {\bibfnamefont {M.}~\bibnamefont {Winstel}},\ }\bibfield  {title}
  {\bibinfo {title} {{Regulator dependence of inhomogeneous phases in the
  (2+1)-dimensional Gross-Neveu model}},\ }\href
  {https://doi.org/10.1103/PhysRevD.103.034503} {\bibfield  {journal} {\bibinfo
   {journal} {Phys. Rev. D}\ }\textbf {\bibinfo {volume} {103}},\ \bibinfo
  {pages} {034503} (\bibinfo {year} {2021})},\ \Eprint
  {https://arxiv.org/abs/2012.09588} {arXiv:2012.09588 [hep-lat]} \BibitemShut
  {NoStop}%
\bibitem [{\citenamefont {Braun}\ \emph {et~al.}(2019)\citenamefont {Braun},
  \citenamefont {Leonhardt},\ and\ \citenamefont {Pawlowski}}]{Braun:2018svj}%
  \BibitemOpen
  \bibfield  {author} {\bibinfo {author} {\bibfnamefont {J.}~\bibnamefont
  {Braun}}, \bibinfo {author} {\bibfnamefont {M.}~\bibnamefont {Leonhardt}},\
  and\ \bibinfo {author} {\bibfnamefont {J.~M.}\ \bibnamefont {Pawlowski}},\
  }\bibfield  {title} {\bibinfo {title} {{Renormalization group consistency and
  low-energy effective theories}},\ }\href
  {https://doi.org/10.21468/SciPostPhys.6.5.056} {\bibfield  {journal}
  {\bibinfo  {journal} {SciPost Phys.}\ }\textbf {\bibinfo {volume} {6}},\
  \bibinfo {pages} {056} (\bibinfo {year} {2019})},\ \Eprint
  {https://arxiv.org/abs/1806.04432} {arXiv:1806.04432 [hep-ph]} \BibitemShut
  {NoStop}%
\bibitem [{\citenamefont {Gell-Mann}\ and\ \citenamefont
  {Levy}(1960)}]{Gell-Mann:1960mvl}%
  \BibitemOpen
  \bibfield  {author} {\bibinfo {author} {\bibfnamefont {M.}~\bibnamefont
  {Gell-Mann}}\ and\ \bibinfo {author} {\bibfnamefont {M.}~\bibnamefont
  {Levy}},\ }\bibfield  {title} {\bibinfo {title} {{The axial vector current in
  beta decay}},\ }\href {https://doi.org/10.1007/BF02859738} {\bibfield
  {journal} {\bibinfo  {journal} {Nuovo Cim.}\ }\textbf {\bibinfo {volume}
  {16}},\ \bibinfo {pages} {705} (\bibinfo {year} {1960})}\BibitemShut
  {NoStop}%
\bibitem [{\citenamefont {Jungnickel}\ and\ \citenamefont
  {Wetterich}(1996)}]{Jungnickel:1995fp}%
  \BibitemOpen
  \bibfield  {author} {\bibinfo {author} {\bibfnamefont {D.~U.}\ \bibnamefont
  {Jungnickel}}\ and\ \bibinfo {author} {\bibfnamefont {C.}~\bibnamefont
  {Wetterich}},\ }\bibfield  {title} {\bibinfo {title} {{Effective action for
  the chiral quark-meson model}},\ }\href
  {https://doi.org/10.1103/PhysRevD.53.5142} {\bibfield  {journal} {\bibinfo
  {journal} {Phys. Rev. D}\ }\textbf {\bibinfo {volume} {53}},\ \bibinfo
  {pages} {5142} (\bibinfo {year} {1996})},\ \Eprint
  {https://arxiv.org/abs/hep-ph/9505267} {arXiv:hep-ph/9505267} \BibitemShut
  {NoStop}%
\bibitem [{\citenamefont {Schaefer}\ and\ \citenamefont
  {Wambach}(2005)}]{Schaefer:2004en}%
  \BibitemOpen
  \bibfield  {author} {\bibinfo {author} {\bibfnamefont {B.-J.}\ \bibnamefont
  {Schaefer}}\ and\ \bibinfo {author} {\bibfnamefont {J.}~\bibnamefont
  {Wambach}},\ }\bibfield  {title} {\bibinfo {title} {{The Phase diagram of the
  quark meson model}},\ }\href
  {https://doi.org/10.1016/j.nuclphysa.2005.04.012} {\bibfield  {journal}
  {\bibinfo  {journal} {Nucl. Phys. A}\ }\textbf {\bibinfo {volume} {757}},\
  \bibinfo {pages} {479} (\bibinfo {year} {2005})},\ \Eprint
  {https://arxiv.org/abs/nucl-th/0403039} {arXiv:nucl-th/0403039} \BibitemShut
  {NoStop}%
\bibitem [{\citenamefont {Tripolt}\ \emph {et~al.}(2014)\citenamefont
  {Tripolt}, \citenamefont {Strodthoff}, \citenamefont {von Smekal},\ and\
  \citenamefont {Wambach}}]{Tripolt:2013jra}%
  \BibitemOpen
  \bibfield  {author} {\bibinfo {author} {\bibfnamefont {R.-A.}\ \bibnamefont
  {Tripolt}}, \bibinfo {author} {\bibfnamefont {N.}~\bibnamefont {Strodthoff}},
  \bibinfo {author} {\bibfnamefont {L.}~\bibnamefont {von Smekal}},\ and\
  \bibinfo {author} {\bibfnamefont {J.}~\bibnamefont {Wambach}},\ }\bibfield
  {title} {\bibinfo {title} {{Spectral Functions for the Quark-Meson Model
  Phase Diagram from the Functional Renormalization Group}},\ }\href
  {https://doi.org/10.1103/PhysRevD.89.034010} {\bibfield  {journal} {\bibinfo
  {journal} {Phys. Rev. D}\ }\textbf {\bibinfo {volume} {89}},\ \bibinfo
  {pages} {034010} (\bibinfo {year} {2014})},\ \Eprint
  {https://arxiv.org/abs/1311.0630} {arXiv:1311.0630 [hep-ph]} \BibitemShut
  {NoStop}%
\bibitem [{\citenamefont {Pawlowski}\ and\ \citenamefont
  {Rennecke}(2014)}]{Pawlowski:2014zaa}%
  \BibitemOpen
  \bibfield  {author} {\bibinfo {author} {\bibfnamefont {J.~M.}\ \bibnamefont
  {Pawlowski}}\ and\ \bibinfo {author} {\bibfnamefont {F.}~\bibnamefont
  {Rennecke}},\ }\bibfield  {title} {\bibinfo {title} {{Higher order
  quark-mesonic scattering processes and the phase structure of QCD}},\ }\href
  {https://doi.org/10.1103/PhysRevD.90.076002} {\bibfield  {journal} {\bibinfo
  {journal} {Phys. Rev.}\ }\textbf {\bibinfo {volume} {D90}},\ \bibinfo {pages}
  {076002} (\bibinfo {year} {2014})},\ \Eprint
  {https://arxiv.org/abs/1403.1179} {arXiv:1403.1179 [hep-ph]} \BibitemShut
  {NoStop}%
\bibitem [{\citenamefont {Kov\'acs}\ \emph {et~al.}(2016)\citenamefont
  {Kov\'acs}, \citenamefont {Sz\'ep},\ and\ \citenamefont
  {Wolf}}]{Kovacs:2016juc}%
  \BibitemOpen
  \bibfield  {author} {\bibinfo {author} {\bibfnamefont {P.}~\bibnamefont
  {Kov\'acs}}, \bibinfo {author} {\bibfnamefont {Z.}~\bibnamefont {Sz\'ep}},\
  and\ \bibinfo {author} {\bibfnamefont {G.}~\bibnamefont {Wolf}},\ }\bibfield
  {title} {\bibinfo {title} {{Existence of the critical endpoint in the vector
  meson extended linear sigma model}},\ }\href
  {https://doi.org/10.1103/PhysRevD.93.114014} {\bibfield  {journal} {\bibinfo
  {journal} {Phys. Rev. D}\ }\textbf {\bibinfo {volume} {93}},\ \bibinfo
  {pages} {114014} (\bibinfo {year} {2016})},\ \Eprint
  {https://arxiv.org/abs/1601.05291} {arXiv:1601.05291 [hep-ph]} \BibitemShut
  {NoStop}%
\bibitem [{\citenamefont {Braun}\ \emph {et~al.}(2016)\citenamefont {Braun},
  \citenamefont {Fister}, \citenamefont {Pawlowski},\ and\ \citenamefont
  {Rennecke}}]{Braun:2014ata}%
  \BibitemOpen
  \bibfield  {author} {\bibinfo {author} {\bibfnamefont {J.}~\bibnamefont
  {Braun}}, \bibinfo {author} {\bibfnamefont {L.}~\bibnamefont {Fister}},
  \bibinfo {author} {\bibfnamefont {J.~M.}\ \bibnamefont {Pawlowski}},\ and\
  \bibinfo {author} {\bibfnamefont {F.}~\bibnamefont {Rennecke}},\ }\bibfield
  {title} {\bibinfo {title} {{From Quarks and Gluons to Hadrons: Chiral
  Symmetry Breaking in Dynamical QCD}},\ }\href
  {https://doi.org/10.1103/PhysRevD.94.034016} {\bibfield  {journal} {\bibinfo
  {journal} {Phys. Rev.}\ }\textbf {\bibinfo {volume} {D94}},\ \bibinfo {pages}
  {034016} (\bibinfo {year} {2016})},\ \Eprint
  {https://arxiv.org/abs/1412.1045} {arXiv:1412.1045 [hep-ph]} \BibitemShut
  {NoStop}%
\bibitem [{\citenamefont {Rennecke}(2015{\natexlab{a}})}]{Rennecke:2015eba}%
  \BibitemOpen
  \bibfield  {author} {\bibinfo {author} {\bibfnamefont {F.}~\bibnamefont
  {Rennecke}},\ }\bibfield  {title} {\bibinfo {title} {{Vacuum structure of
  vector mesons in QCD}},\ }\href {https://doi.org/10.1103/PhysRevD.92.076012}
  {\bibfield  {journal} {\bibinfo  {journal} {Phys. Rev.}\ }\textbf {\bibinfo
  {volume} {D92}},\ \bibinfo {pages} {076012} (\bibinfo {year}
  {2015}{\natexlab{a}})},\ \Eprint {https://arxiv.org/abs/1504.03585}
  {arXiv:1504.03585 [hep-ph]} \BibitemShut {NoStop}%
\bibitem [{\citenamefont {Rennecke}(2015{\natexlab{b}})}]{Rennecke:2015lur}%
  \BibitemOpen
  \bibfield  {author} {\bibinfo {author} {\bibfnamefont {F.}~\bibnamefont
  {Rennecke}},\ }\emph {\bibinfo {title} {{The Chiral Phase Transition of
  QCD.}}},\ \href {https://doi.org/10.11588/heidok.00019205} {Ph.D. thesis},\
  \bibinfo  {school} {U. Heidelberg (main)} (\bibinfo {year}
  {2015}{\natexlab{b}})\BibitemShut {NoStop}%
\bibitem [{\citenamefont {Cyrol}\ \emph {et~al.}(2018)\citenamefont {Cyrol},
  \citenamefont {Mitter}, \citenamefont {Pawlowski},\ and\ \citenamefont
  {Strodthoff}}]{Cyrol:2017ewj}%
  \BibitemOpen
  \bibfield  {author} {\bibinfo {author} {\bibfnamefont {A.~K.}\ \bibnamefont
  {Cyrol}}, \bibinfo {author} {\bibfnamefont {M.}~\bibnamefont {Mitter}},
  \bibinfo {author} {\bibfnamefont {J.~M.}\ \bibnamefont {Pawlowski}},\ and\
  \bibinfo {author} {\bibfnamefont {N.}~\bibnamefont {Strodthoff}},\ }\bibfield
   {title} {\bibinfo {title} {{Nonperturbative quark, gluon, and meson
  correlators of unquenched QCD}},\ }\href
  {https://doi.org/10.1103/PhysRevD.97.054006} {\bibfield  {journal} {\bibinfo
  {journal} {Phys. Rev.}\ }\textbf {\bibinfo {volume} {D97}},\ \bibinfo {pages}
  {054006} (\bibinfo {year} {2018})},\ \Eprint
  {https://arxiv.org/abs/1706.06326} {arXiv:1706.06326 [hep-ph]} \BibitemShut
  {NoStop}%
\bibitem [{\citenamefont {Ihssen}\ \emph {et~al.}(2024)\citenamefont {Ihssen},
  \citenamefont {Pawlowski}, \citenamefont {Sattler},\ and\ \citenamefont
  {Wink}}]{Ihssen:2024miv}%
  \BibitemOpen
  \bibfield  {author} {\bibinfo {author} {\bibfnamefont {F.}~\bibnamefont
  {Ihssen}}, \bibinfo {author} {\bibfnamefont {J.~M.}\ \bibnamefont
  {Pawlowski}}, \bibinfo {author} {\bibfnamefont {F.~R.}\ \bibnamefont
  {Sattler}},\ and\ \bibinfo {author} {\bibfnamefont {N.}~\bibnamefont
  {Wink}},\ }\bibfield  {title} {\bibinfo {title} {{Towards quantitative
  precision in functional QCD I}},\ }\href@noop {} {\  (\bibinfo {year}
  {2024})},\ \Eprint {https://arxiv.org/abs/2408.08413} {arXiv:2408.08413
  [hep-ph]} \BibitemShut {NoStop}%
\bibitem [{\citenamefont {Brandt}\ \emph {et~al.}(2025)\citenamefont {Brandt},
  \citenamefont {Chelnokov}, \citenamefont {Endrodi}, \citenamefont {Marko},
  \citenamefont {Scheid},\ and\ \citenamefont {von Smekal}}]{Brandt:2025tkg}%
  \BibitemOpen
  \bibfield  {author} {\bibinfo {author} {\bibfnamefont {B.~B.}\ \bibnamefont
  {Brandt}}, \bibinfo {author} {\bibfnamefont {V.}~\bibnamefont {Chelnokov}},
  \bibinfo {author} {\bibfnamefont {G.}~\bibnamefont {Endrodi}}, \bibinfo
  {author} {\bibfnamefont {G.}~\bibnamefont {Marko}}, \bibinfo {author}
  {\bibfnamefont {D.}~\bibnamefont {Scheid}},\ and\ \bibinfo {author}
  {\bibfnamefont {L.}~\bibnamefont {von Smekal}},\ }\bibfield  {title}
  {\bibinfo {title} {{Renormalization group invariant mean-field model for QCD
  at finite isospin density}},\ }\href {https://doi.org/10.1103/fryz-f3vw}
  {\bibfield  {journal} {\bibinfo  {journal} {Phys. Rev. D}\ }\textbf {\bibinfo
  {volume} {112}},\ \bibinfo {pages} {054038} (\bibinfo {year} {2025})},\
  \Eprint {https://arxiv.org/abs/2502.04025} {arXiv:2502.04025 [hep-ph]}
  \BibitemShut {NoStop}%
\bibitem [{\citenamefont {Gholami}\ \emph {et~al.}(2025)\citenamefont
  {Gholami}, \citenamefont {Kurth}, \citenamefont {Mire}, \citenamefont
  {Buballa},\ and\ \citenamefont {Schaefer}}]{Gholami:2025afm}%
  \BibitemOpen
  \bibfield  {author} {\bibinfo {author} {\bibfnamefont {H.}~\bibnamefont
  {Gholami}}, \bibinfo {author} {\bibfnamefont {L.}~\bibnamefont {Kurth}},
  \bibinfo {author} {\bibfnamefont {U.}~\bibnamefont {Mire}}, \bibinfo {author}
  {\bibfnamefont {M.}~\bibnamefont {Buballa}},\ and\ \bibinfo {author}
  {\bibfnamefont {B.-J.}\ \bibnamefont {Schaefer}},\ }\bibfield  {title}
  {\bibinfo {title} {{Renormalizing the Quark-Meson-Diquark Model}},\
  }\href@noop {} {\  (\bibinfo {year} {2025})},\ \Eprint
  {https://arxiv.org/abs/2505.22542} {arXiv:2505.22542 [hep-ph]} \BibitemShut
  {NoStop}%
\bibitem [{\citenamefont {Carignano}\ \emph {et~al.}(2014)\citenamefont
  {Carignano}, \citenamefont {Buballa},\ and\ \citenamefont
  {Schaefer}}]{Carignano:2014jla}%
  \BibitemOpen
  \bibfield  {author} {\bibinfo {author} {\bibfnamefont {S.}~\bibnamefont
  {Carignano}}, \bibinfo {author} {\bibfnamefont {M.}~\bibnamefont {Buballa}},\
  and\ \bibinfo {author} {\bibfnamefont {B.-J.}\ \bibnamefont {Schaefer}},\
  }\bibfield  {title} {\bibinfo {title} {{Inhomogeneous phases in the
  quark-meson model with vacuum fluctuations}},\ }\href
  {https://doi.org/10.1103/PhysRevD.90.014033} {\bibfield  {journal} {\bibinfo
  {journal} {Phys. Rev.}\ }\textbf {\bibinfo {volume} {D90}},\ \bibinfo {pages}
  {014033} (\bibinfo {year} {2014})},\ \Eprint
  {https://arxiv.org/abs/1404.0057} {arXiv:1404.0057 [hep-ph]} \BibitemShut
  {NoStop}%
\bibitem [{\citenamefont {Carignano}\ \emph {et~al.}(2016)\citenamefont
  {Carignano}, \citenamefont {Buballa},\ and\ \citenamefont
  {Elkamhawy}}]{Carignano:2016jnw}%
  \BibitemOpen
  \bibfield  {author} {\bibinfo {author} {\bibfnamefont {S.}~\bibnamefont
  {Carignano}}, \bibinfo {author} {\bibfnamefont {M.}~\bibnamefont {Buballa}},\
  and\ \bibinfo {author} {\bibfnamefont {W.}~\bibnamefont {Elkamhawy}},\
  }\bibfield  {title} {\bibinfo {title} {{Consistent parameter fixing in the
  quark-meson model with vacuum fluctuations}},\ }\href
  {https://doi.org/10.1103/PhysRevD.94.034023} {\bibfield  {journal} {\bibinfo
  {journal} {Phys. Rev. D}\ }\textbf {\bibinfo {volume} {94}},\ \bibinfo
  {pages} {034023} (\bibinfo {year} {2016})},\ \Eprint
  {https://arxiv.org/abs/1606.08859} {arXiv:1606.08859 [hep-ph]} \BibitemShut
  {NoStop}%
\bibitem [{\citenamefont {Adhikari}\ \emph
  {et~al.}(2017{\natexlab{a}})\citenamefont {Adhikari}, \citenamefont
  {Andersen},\ and\ \citenamefont {Kneschke}}]{Adhikari:2016eef}%
  \BibitemOpen
  \bibfield  {author} {\bibinfo {author} {\bibfnamefont {P.}~\bibnamefont
  {Adhikari}}, \bibinfo {author} {\bibfnamefont {J.~O.}\ \bibnamefont
  {Andersen}},\ and\ \bibinfo {author} {\bibfnamefont {P.}~\bibnamefont
  {Kneschke}},\ }\bibfield  {title} {\bibinfo {title} {{On-shell parameter
  fixing in the quark-meson model}},\ }\href
  {https://doi.org/10.1103/PhysRevD.95.036017} {\bibfield  {journal} {\bibinfo
  {journal} {Phys. Rev. D}\ }\textbf {\bibinfo {volume} {95}},\ \bibinfo
  {pages} {036017} (\bibinfo {year} {2017}{\natexlab{a}})},\ \Eprint
  {https://arxiv.org/abs/1612.03668} {arXiv:1612.03668 [hep-ph]} \BibitemShut
  {NoStop}%
\bibitem [{\citenamefont {Adhikari}\ \emph
  {et~al.}(2017{\natexlab{b}})\citenamefont {Adhikari}, \citenamefont
  {Andersen},\ and\ \citenamefont {Kneschke}}]{Adhikari:2017ydi}%
  \BibitemOpen
  \bibfield  {author} {\bibinfo {author} {\bibfnamefont {P.}~\bibnamefont
  {Adhikari}}, \bibinfo {author} {\bibfnamefont {J.~O.}\ \bibnamefont
  {Andersen}},\ and\ \bibinfo {author} {\bibfnamefont {P.}~\bibnamefont
  {Kneschke}},\ }\bibfield  {title} {\bibinfo {title} {{Inhomogeneous chiral
  condensate in the quark-meson model}},\ }\href
  {https://doi.org/10.1103/PhysRevD.96.016013} {\bibfield  {journal} {\bibinfo
  {journal} {Phys. Rev. D}\ }\textbf {\bibinfo {volume} {96}},\ \bibinfo
  {pages} {016013} (\bibinfo {year} {2017}{\natexlab{b}})},\ \bibinfo {note}
  {[Erratum: Phys.Rev.D 98, 099902 (2018)]},\ \Eprint
  {https://arxiv.org/abs/1702.01324} {arXiv:1702.01324 [hep-ph]} \BibitemShut
  {NoStop}%
\bibitem [{\citenamefont {Buballa}\ \emph {et~al.}(2020)\citenamefont
  {Buballa}, \citenamefont {Carignano},\ and\ \citenamefont
  {Kurth}}]{Buballa:2020xaa}%
  \BibitemOpen
  \bibfield  {author} {\bibinfo {author} {\bibfnamefont {M.}~\bibnamefont
  {Buballa}}, \bibinfo {author} {\bibfnamefont {S.}~\bibnamefont {Carignano}},\
  and\ \bibinfo {author} {\bibfnamefont {L.}~\bibnamefont {Kurth}},\ }\bibfield
   {title} {\bibinfo {title} {{Inhomogeneous phases in the quark-meson model
  with explicit chiral-symmetry breaking}},\ }\href
  {https://doi.org/10.1140/epjst/e2020-000101-x} {\bibfield  {journal}
  {\bibinfo  {journal} {Eur. Phys. J. ST}\ }\textbf {\bibinfo {volume} {229}},\
  \bibinfo {pages} {3371} (\bibinfo {year} {2020})},\ \Eprint
  {https://arxiv.org/abs/2006.02133} {arXiv:2006.02133 [hep-ph]} \BibitemShut
  {NoStop}%
\bibitem [{\citenamefont {Rai}\ and\ \citenamefont
  {Tiwari}(2022)}]{Rai:2022wth}%
  \BibitemOpen
  \bibfield  {author} {\bibinfo {author} {\bibfnamefont {S.~K.}\ \bibnamefont
  {Rai}}\ and\ \bibinfo {author} {\bibfnamefont {V.~K.}\ \bibnamefont
  {Tiwari}},\ }\bibfield  {title} {\bibinfo {title} {{On-shell versus curvature
  mass parameter fixing schemes in the quark-meson model and its phase
  diagrams}},\ }\href {https://doi.org/10.1103/PhysRevD.105.094010} {\bibfield
  {journal} {\bibinfo  {journal} {Phys. Rev. D}\ }\textbf {\bibinfo {volume}
  {105}},\ \bibinfo {pages} {094010} (\bibinfo {year} {2022})},\ \Eprint
  {https://arxiv.org/abs/2202.11661} {arXiv:2202.11661 [hep-ph]} \BibitemShut
  {NoStop}%
\bibitem [{\citenamefont {Yin}\ \emph {et~al.}(2019)\citenamefont {Yin},
  \citenamefont {Wen},\ and\ \citenamefont {Fu}}]{Yin:2019ebz}%
  \BibitemOpen
  \bibfield  {author} {\bibinfo {author} {\bibfnamefont {S.}~\bibnamefont
  {Yin}}, \bibinfo {author} {\bibfnamefont {R.}~\bibnamefont {Wen}},\ and\
  \bibinfo {author} {\bibfnamefont {W.-j.}\ \bibnamefont {Fu}},\ }\bibfield
  {title} {\bibinfo {title} {{Mesonic dynamics and the QCD phase transition}},\
  }\href {https://doi.org/10.1103/PhysRevD.100.094029} {\bibfield  {journal}
  {\bibinfo  {journal} {Phys. Rev. D}\ }\textbf {\bibinfo {volume} {100}},\
  \bibinfo {pages} {094029} (\bibinfo {year} {2019})},\ \Eprint
  {https://arxiv.org/abs/1907.10262} {arXiv:1907.10262 [hep-ph]} \BibitemShut
  {NoStop}%
\bibitem [{\citenamefont {Skokov}\ \emph {et~al.}(2010)\citenamefont {Skokov},
  \citenamefont {Friman}, \citenamefont {Nakano}, \citenamefont {Redlich},\
  and\ \citenamefont {Schaefer}}]{Skokov:2010sf}%
  \BibitemOpen
  \bibfield  {author} {\bibinfo {author} {\bibfnamefont {V.}~\bibnamefont
  {Skokov}}, \bibinfo {author} {\bibfnamefont {B.}~\bibnamefont {Friman}},
  \bibinfo {author} {\bibfnamefont {E.}~\bibnamefont {Nakano}}, \bibinfo
  {author} {\bibfnamefont {K.}~\bibnamefont {Redlich}},\ and\ \bibinfo {author}
  {\bibfnamefont {B.~J.}\ \bibnamefont {Schaefer}},\ }\bibfield  {title}
  {\bibinfo {title} {{Vacuum fluctuations and the thermodynamics of chiral
  models}},\ }\href {https://doi.org/10.1103/PhysRevD.82.034029} {\bibfield
  {journal} {\bibinfo  {journal} {Phys. Rev. D}\ }\textbf {\bibinfo {volume}
  {82}},\ \bibinfo {pages} {034029} (\bibinfo {year} {2010})},\ \Eprint
  {https://arxiv.org/abs/1005.3166} {arXiv:1005.3166 [hep-ph]} \BibitemShut
  {NoStop}%
\bibitem [{\citenamefont {Mukherjee}\ \emph {et~al.}(2022)\citenamefont
  {Mukherjee}, \citenamefont {Rennecke},\ and\ \citenamefont
  {Skokov}}]{Mukherjee:2021tyg}%
  \BibitemOpen
  \bibfield  {author} {\bibinfo {author} {\bibfnamefont {S.}~\bibnamefont
  {Mukherjee}}, \bibinfo {author} {\bibfnamefont {F.}~\bibnamefont
  {Rennecke}},\ and\ \bibinfo {author} {\bibfnamefont {V.~V.}\ \bibnamefont
  {Skokov}},\ }\bibfield  {title} {\bibinfo {title} {{Analytical structure of
  the equation of state at finite density: Resummation versus expansion in a
  low energy model}},\ }\href {https://doi.org/10.1103/PhysRevD.105.014026}
  {\bibfield  {journal} {\bibinfo  {journal} {Phys. Rev. D}\ }\textbf {\bibinfo
  {volume} {105}},\ \bibinfo {pages} {014026} (\bibinfo {year} {2022})},\
  \Eprint {https://arxiv.org/abs/2110.02241} {arXiv:2110.02241 [hep-ph]}
  \BibitemShut {NoStop}%
\bibitem [{\citenamefont {Peskin}\ and\ \citenamefont
  {Schroeder}(1995)}]{Peskin:1995ev}%
  \BibitemOpen
  \bibfield  {author} {\bibinfo {author} {\bibfnamefont {M.~E.}\ \bibnamefont
  {Peskin}}\ and\ \bibinfo {author} {\bibfnamefont {D.~V.}\ \bibnamefont
  {Schroeder}},\ }\href {https://doi.org/10.1201/9780429503559} {\emph
  {\bibinfo {title} {{An Introduction to quantum field theory}}}}\ (\bibinfo
  {publisher} {Addison-Wesley},\ \bibinfo {address} {Reading, USA},\ \bibinfo
  {year} {1995})\BibitemShut {NoStop}%
\bibitem [{Note2()}]{Note2}%
  \BibitemOpen
  \bibinfo {note} {While there is no explicit renormalization condition for the
  wave function renormalization in Refs.\ \cite {Carignano:2014jla,
  Carignano:2016jnw, Buballa:2020xaa}, $Z_\pi $ is implicitly fixed through
  $f_\pi $. The advantage of an explicit condition for the wave function
  renormalization is that it straightforwardly generalizes also to other
  mesons.}\BibitemShut {Stop}%
\bibitem [{\citenamefont {Bazavov}\ \emph
  {et~al.}(2019{\natexlab{b}})\citenamefont {Bazavov} \emph
  {et~al.}}]{Bazavov:2019www}%
  \BibitemOpen
  \bibfield  {author} {\bibinfo {author} {\bibfnamefont {A.}~\bibnamefont
  {Bazavov}} \emph {et~al.},\ }\bibfield  {title} {\bibinfo {title} {{Meson
  screening masses in (2+1)-flavor QCD}},\ }\href
  {https://doi.org/10.1103/PhysRevD.100.094510} {\bibfield  {journal} {\bibinfo
   {journal} {Phys. Rev. D}\ }\textbf {\bibinfo {volume} {100}},\ \bibinfo
  {pages} {094510} (\bibinfo {year} {2019}{\natexlab{b}})},\ \Eprint
  {https://arxiv.org/abs/1908.09552} {arXiv:1908.09552 [hep-lat]} \BibitemShut
  {NoStop}%
\bibitem [{Note3()}]{Note3}%
  \BibitemOpen
  \bibinfo {note} {Note that $Z_\pi ^\parallel = Z_\pi ^\perp $ in vacuum, and
  this remains a good approximation even at moderate $T$ and $\mu $ \cite
  {Yin:2019ebz}. However, it clearly breaks down when the system enters the
  moat regime, as $Z^\perp $ changes sign while $Z^\parallel $ is always
  positive due to causality.}\BibitemShut {Stop}%
\bibitem [{Note4()}]{Note4}%
  \BibitemOpen
  \bibinfo {note} {We note that a similar result could presumably be achieved
  by using some form of Pauli-Villars regularization}\BibitemShut {NoStop}%
\bibitem [{\citenamefont {Rennecke}\ and\ \citenamefont {Yin}(2025)}]{DM2}%
  \BibitemOpen
  \bibfield  {author} {\bibinfo {author} {\bibfnamefont {F.}~\bibnamefont
  {Rennecke}}\ and\ \bibinfo {author} {\bibfnamefont {S.}~\bibnamefont {Yin}},\
  }\bibfield  {title} {\bibinfo {title} {{Dissecting the moat regime at low
  energies II: Correlations}},\ }\href@noop {} {\bibfield  {journal} {\bibinfo
  {journal} {in preparation}\ } (\bibinfo {year} {2025})}\BibitemShut {NoStop}%
\bibitem [{Note5()}]{Note5}%
  \BibitemOpen
  \bibinfo {note} {Note that the solution of the gap equation only depends on
  the curvature masses in RPA, so the chiral phase boundary is not affected by
  changes in $Z$}\BibitemShut {NoStop}%
\bibitem [{\citenamefont {Nickel}(2009)}]{Nickel:2009ke}%
  \BibitemOpen
  \bibfield  {author} {\bibinfo {author} {\bibfnamefont {D.}~\bibnamefont
  {Nickel}},\ }\bibfield  {title} {\bibinfo {title} {{How many phases meet at
  the chiral critical point?}},\ }\href
  {https://doi.org/10.1103/PhysRevLett.103.072301} {\bibfield  {journal}
  {\bibinfo  {journal} {Phys. Rev. Lett.}\ }\textbf {\bibinfo {volume} {103}},\
  \bibinfo {pages} {072301} (\bibinfo {year} {2009})},\ \Eprint
  {https://arxiv.org/abs/0902.1778} {arXiv:0902.1778 [hep-ph]} \BibitemShut
  {NoStop}%
\bibitem [{Note6()}]{Note6}%
  \BibitemOpen
  \bibinfo {note} {We thank Michael Buballa for pointing this out.}\BibitemShut
  {Stop}%
\bibitem [{\citenamefont {Klevansky}(1992)}]{Klevansky:1992qe}%
  \BibitemOpen
  \bibfield  {author} {\bibinfo {author} {\bibfnamefont {S.~P.}\ \bibnamefont
  {Klevansky}},\ }\bibfield  {title} {\bibinfo {title} {{The Nambu-Jona-Lasinio
  model of quantum chromodynamics}},\ }\href
  {https://doi.org/10.1103/RevModPhys.64.649} {\bibfield  {journal} {\bibinfo
  {journal} {Rev. Mod. Phys.}\ }\textbf {\bibinfo {volume} {64}},\ \bibinfo
  {pages} {649} (\bibinfo {year} {1992})}\BibitemShut {NoStop}%
\bibitem [{\citenamefont {fQCD collaboration}()}]{fQCD}%
  \BibitemOpen
  \bibfield  {author} {\bibinfo {author} {\bibnamefont {fQCD collaboration}},\
  }\href {https://fqcd-collaboration.github.io} {\bibinfo  {journal}
  {https://fqcd-collaboration.github.io}\ }\BibitemShut {NoStop}%
\bibitem [{\citenamefont {Cohen}(2003)}]{Cohen:2003kd}%
  \BibitemOpen
\bibfield  {journal} {  }\bibfield  {author} {\bibinfo {author} {\bibfnamefont
  {T.~D.}\ \bibnamefont {Cohen}},\ }\bibfield  {title} {\bibinfo {title}
  {{Functional integrals for QCD at nonzero chemical potential and zero
  density}},\ }\href {https://doi.org/10.1103/PhysRevLett.91.222001} {\bibfield
   {journal} {\bibinfo  {journal} {Phys. Rev. Lett.}\ }\textbf {\bibinfo
  {volume} {91}},\ \bibinfo {pages} {222001} (\bibinfo {year} {2003})},\
  \Eprint {https://arxiv.org/abs/hep-ph/0307089} {arXiv:hep-ph/0307089}
  \BibitemShut {NoStop}%
\bibitem [{\citenamefont {Mark{\'o}}\ \emph {et~al.}(2014)\citenamefont
  {Mark{\'o}}, \citenamefont {Reinosa},\ and\ \citenamefont
  {Sz{\'e}p}}]{Marko:2014hea}%
  \BibitemOpen
  \bibfield  {author} {\bibinfo {author} {\bibfnamefont {G.}~\bibnamefont
  {Mark{\'o}}}, \bibinfo {author} {\bibfnamefont {U.}~\bibnamefont {Reinosa}},\
  and\ \bibinfo {author} {\bibfnamefont {Z.}~\bibnamefont {Sz{\'e}p}},\
  }\bibfield  {title} {\bibinfo {title} {{Bose-Einstein condensation and Silver
  Blaze property from the two-loop $\Phi$-derivable approximation}},\ }\href
  {https://doi.org/10.1103/PhysRevD.90.125021} {\bibfield  {journal} {\bibinfo
  {journal} {Phys. Rev. D}\ }\textbf {\bibinfo {volume} {90}},\ \bibinfo
  {pages} {125021} (\bibinfo {year} {2014})},\ \Eprint
  {https://arxiv.org/abs/1410.6998} {arXiv:1410.6998 [hep-ph]} \BibitemShut
  {NoStop}%
\bibitem [{\citenamefont {Khan}\ \emph {et~al.}(2015)\citenamefont {Khan},
  \citenamefont {Pawlowski}, \citenamefont {Rennecke},\ and\ \citenamefont
  {Scherer}}]{Khan:2015puu}%
  \BibitemOpen
  \bibfield  {author} {\bibinfo {author} {\bibfnamefont {N.}~\bibnamefont
  {Khan}}, \bibinfo {author} {\bibfnamefont {J.~M.}\ \bibnamefont {Pawlowski}},
  \bibinfo {author} {\bibfnamefont {F.}~\bibnamefont {Rennecke}},\ and\
  \bibinfo {author} {\bibfnamefont {M.~M.}\ \bibnamefont {Scherer}},\
  }\bibfield  {title} {\bibinfo {title} {{The Phase Diagram of QC2D from
  Functional Methods}},\ }\href@noop {} {\  (\bibinfo {year} {2015})},\ \Eprint
  {https://arxiv.org/abs/1512.03673} {arXiv:1512.03673 [hep-ph]} \BibitemShut
  {NoStop}%
\bibitem [{\citenamefont {Fu}\ \emph {et~al.}(2016)\citenamefont {Fu},
  \citenamefont {Pawlowski}, \citenamefont {Rennecke},\ and\ \citenamefont
  {Schaefer}}]{Fu:2016tey}%
  \BibitemOpen
  \bibfield  {author} {\bibinfo {author} {\bibfnamefont {W.-j.}\ \bibnamefont
  {Fu}}, \bibinfo {author} {\bibfnamefont {J.~M.}\ \bibnamefont {Pawlowski}},
  \bibinfo {author} {\bibfnamefont {F.}~\bibnamefont {Rennecke}},\ and\
  \bibinfo {author} {\bibfnamefont {B.-J.}\ \bibnamefont {Schaefer}},\
  }\bibfield  {title} {\bibinfo {title} {{Baryon number fluctuations at finite
  temperature and density}},\ }\href
  {https://doi.org/10.1103/PhysRevD.94.116020} {\bibfield  {journal} {\bibinfo
  {journal} {Phys. Rev. D}\ }\textbf {\bibinfo {volume} {94}},\ \bibinfo
  {pages} {116020} (\bibinfo {year} {2016})},\ \Eprint
  {https://arxiv.org/abs/1608.04302} {arXiv:1608.04302 [hep-ph]} \BibitemShut
  {NoStop}%
\bibitem [{\citenamefont {Aboona}\ \emph {et~al.}(2025)\citenamefont {Aboona}
  \emph {et~al.}}]{STAR:2025zdq}%
  \BibitemOpen
  \bibfield  {author} {\bibinfo {author} {\bibfnamefont {B.~E.}\ \bibnamefont
  {Aboona}} \emph {et~al.} (\bibinfo {collaboration} {STAR}),\ }\bibfield
  {title} {\bibinfo {title} {{Precision Measurement of Net-Proton-Number
  Fluctuations in Au+Au Collisions at RHIC}},\ }\href
  {https://doi.org/10.1103/9l69-2d7p} {\bibfield  {journal} {\bibinfo
  {journal} {Phys. Rev. Lett.}\ }\textbf {\bibinfo {volume} {135}},\ \bibinfo
  {pages} {142301} (\bibinfo {year} {2025})},\ \Eprint
  {https://arxiv.org/abs/2504.00817} {arXiv:2504.00817 [nucl-ex]} \BibitemShut
  {NoStop}%
\bibitem [{\citenamefont {Bazavov}\ \emph {et~al.}(2017)\citenamefont {Bazavov}
  \emph {et~al.}}]{Bazavov:2017dus}%
  \BibitemOpen
  \bibfield  {author} {\bibinfo {author} {\bibfnamefont {A.}~\bibnamefont
  {Bazavov}} \emph {et~al.},\ }\bibfield  {title} {\bibinfo {title} {{The QCD
  Equation of State to $\mathcal{O}(\mu_B^6)$ from Lattice QCD}},\ }\href
  {https://doi.org/10.1103/PhysRevD.95.054504} {\bibfield  {journal} {\bibinfo
  {journal} {Phys. Rev.}\ }\textbf {\bibinfo {volume} {D95}},\ \bibinfo {pages}
  {054504} (\bibinfo {year} {2017})},\ \Eprint
  {https://arxiv.org/abs/1701.04325} {arXiv:1701.04325 [hep-lat]} \BibitemShut
  {NoStop}%
\bibitem [{\citenamefont {Borsanyi}\ \emph {et~al.}(2025)\citenamefont
  {Borsanyi}, \citenamefont {Fodor}, \citenamefont {Guenther}, \citenamefont
  {Parotto}, \citenamefont {Pasztor}, \citenamefont {Ratti}, \citenamefont
  {Vovchenko},\ and\ \citenamefont {Wong}}]{Borsanyi:2025dyp}%
  \BibitemOpen
  \bibfield  {author} {\bibinfo {author} {\bibfnamefont {S.}~\bibnamefont
  {Borsanyi}}, \bibinfo {author} {\bibfnamefont {Z.}~\bibnamefont {Fodor}},
  \bibinfo {author} {\bibfnamefont {J.~N.}\ \bibnamefont {Guenther}}, \bibinfo
  {author} {\bibfnamefont {P.}~\bibnamefont {Parotto}}, \bibinfo {author}
  {\bibfnamefont {A.}~\bibnamefont {Pasztor}}, \bibinfo {author} {\bibfnamefont
  {C.}~\bibnamefont {Ratti}}, \bibinfo {author} {\bibfnamefont
  {V.}~\bibnamefont {Vovchenko}},\ and\ \bibinfo {author} {\bibfnamefont
  {C.~H.}\ \bibnamefont {Wong}},\ }\bibfield  {title} {\bibinfo {title}
  {{Lattice QCD constraints on the critical point from an improved precision
  equation of state}},\ }\href@noop {} {\  (\bibinfo {year} {2025})},\ \Eprint
  {https://arxiv.org/abs/2502.10267} {arXiv:2502.10267 [hep-lat]} \BibitemShut
  {NoStop}%
\end{thebibliography}%

\end{document}